\documentclass[journal]{IEEEtran}

\usepackage{cite}
\usepackage{amssymb,amsmath, amsfonts, amsthm}
\usepackage{xcolor}
\usepackage[caption=false,font=footnotesize]{subfig}
\usepackage{tikz}
\usetikzlibrary{shapes.geometric, decorations.pathmorphing, positioning, arrows, arrows.meta, patterns, shapes, calc, fit, backgrounds}
\usepackage{graphicx}
\graphicspath{{./figures/}}
\DeclareGraphicsExtensions{.pdf,.jpeg,.png,.eps}

\newcommand{\R}{\mathbb R}

\newcommand{\K}{\mathbb K}

\newcommand{\T}{\mathbb T}
\newcommand{\Pbb}{\mathbb P}

\newcommand{\Lbb}{\mathbb L}

\newcommand{\TT}{\mathcal T}
\newcommand{\OO}{\mathcal O}

\newcommand{\KK}{\mathcal K}

\newcommand{\Tfrak}{\mathfrak T}
\newcommand{\Kfrak}{\mathfrak K}
\newcommand{\Zfrak}{\mathfrak Z}

\newcommand{\Qfrak}{\mathfrak Q}
\newcommand{\jfrakbm}{\bm{\mathfrak j}}
\newcommand{\mfrakbm}{\bm{\mathfrak m}}
\newcommand{\ufrakbm}{\bm{\mathfrak u}}
\newcommand{\vfrakbm}{\bm{\mathfrak v}}

\newcommand{\efrakbm}{\bm{\mathfrak e}}
\newcommand{\hfrakbm}{\bm{\mathfrak h}}

\DeclareMathOperator{\xxs}{\mathbf x}

\DeclareMathOperator{\ees}{\mathbf e}
\DeclareMathOperator{\hhs}{\mathbf h}

\DeclareMathOperator{\rrs}{\mathbf r}
\DeclareMathOperator{\jjs}{\mathbf j}

\DeclareMathOperator{\vvs}{\mathbf v}
\DeclareMathOperator{\wws}{\mathbf w}
\DeclareMathOperator{\ms}{\mathbf m}

\DeclareMathOperator{\LLs}{\mathbf L}

\DeclareMathOperator{\Hs}{H}
\DeclareMathOperator{\HHs}{\mathbf H}

\DeclareMathOperator{\KKs}{\mathbf K}

\DeclareMathOperator{\QQs}{\mathbf Q}

\DeclareMathOperator{\XXs}{\mathbf X}

\DeclareMathOperator{\TTs}{\mathbf T}

\DeclareMathOperator{\PPo}{\mathbf P}
\DeclareMathOperator{\MMs}{\mathbf M}
\DeclareMathOperator{\GGs}{\mathbf G}
\DeclareMathOperator{\RRs}{\mathbf R}
\DeclareMathOperator{\ZZs}{\mathbf Z}

\newcommand{\brac}[1]{\left\lbrace{#1}\right\rbrace}

\newcommand{\paren}[1]{\left({#1}\right)}

\newcommand{\abs}[1]{\left\vert{#1}\right\vert}

\newcommand{\inprod}[1]{\left\langle{#1}\right\rangle}

\newcommand{\wt}[1]{\widetilde{#1}}
\newcommand{\wh}[1]{\widehat{#1}}

\newcommand{\transpose}{\mathsf{T}}

\newcommand{\bm}[1]{\boldsymbol{#1}}

\newcommand{\curlt}{\textbf{\textup{curl}}}
\newcommand{\divt}{\textup{div}}
\newcommand{\gradt}{\textbf{grad}}

\newcommand{\eps}{\epsilon}

\newcommand{\pa}{\partial}

\newcommand{\Gm}{\Gamma}
\newcommand{\om}{\omega}

\newcommand{\Sm}{\Sigma}

\newcommand{\xb}{\bm{x}}
\newcommand{\yb}{\bm{y}}

\newcommand{\hb}{\bm{h}}

\newcommand{\eb}{\bm{e}}

\newcommand{\ub}{\bm{u}}

\newcommand{\fb}{\bm{f}}
\newcommand{\gb}{\bm{g}}
\newcommand{\kb}{\bm{k}}
\newcommand{\jb}{\bm{j}}

\newcommand{\mb}{\bm{m}}

\newcommand{\pb}{\bm{p}}

\newcommand{\nv}{\textbf{n}}
\newcommand{\uv}{\textbf{u}}

\newcommand{\zrb}{\bm{0}}

\newcommand{\Lmdb}{\bm{\Lambda}}
\newcommand{\Smb}{\bm{\Sm}}

\newcommand{\ds}{\,\mathrm{d}s}
\newcommand{\dt}{\,\mathrm{d}t}
\newcommand{\di}{\,\mathrm{d}}

\newcommand{\q}{\quad}
\newcommand{\qq}{\qquad}
\newcommand{\qqq}{\qquad\quad}
\newcommand{\qqqq}{\qquad\qquad}

\begin{document}

\title{A Stable, Accurate, and Well-Conditioned Time-Domain PMCHWT Formulation}

\author{Van~Chien~Le,~\IEEEmembership{Member,~IEEE,}
        Cedric~M\"{u}nger,~\IEEEmembership{Member,~IEEE,}
        Francesco~P.~Andriulli,~\IEEEmembership{Fellow,~IEEE,}
        and~Kristof~Cools,~\IEEEmembership{Member,~IEEE}
\thanks{Manuscript received April 19, 2005; revised August 26, 2008; accepted
19 May 2014. Date of publication 12 June 2014; date of current version 9 July
2014. This work was supported by European Research Council (ERC) under European Union's Horizon 2020 Research and Innovation Program under Grant 101001847 and the Special Research Fund (BOF) of Ghent University under Grant BOF.PDO.2024.0016.01. The first author gratefully acknowledges support from the Research Foundation -- Flanders (FWO) for his research stay at the Politecnico di Torino (file ID V425624N), during which this manuscript was prepared. \textit{(Corresponding author: Van Chien Le.)}}%
\thanks{V.~C. Le, C. M\"{u}nger and K. Cools are with the IDLab, Department of Information Technology at Ghent University -- imec, 9000 Ghent, Belgium (e-mail: vanchien.le@ugent.be, cedric.muenger@ugent.be, kristof.cools@ugent.be).}
\thanks{F.~P. Andriulli is with the Department of Electronics and Telecommunications, Politecnico di Torino, 10129 Turin, Italy (e-mail: francesco.andriulli@polito.it).}}


\maketitle
 
\begin{abstract}
    This paper introduces a new boundary element formulation for transient electromagnetic scattering by homogeneous dielectric objects, based on the time-domain PMCHWT equation. To address dense-mesh breakdown, a multiplicative Calder\'{o}n preconditioner constructed from a modified static electric field integral operator is employed. Large-timestep breakdown and late-time instability are simultaneously resolved through a rescaling of the Helmholtz components using quasi-Helmholtz projectors, with temporal differentiation and integration serving as the rescaling operators. This rescaling additionally balances the loop and star components in the large-timestep regime, thereby preventing loss of accuracy in the secondary quantities caused by numerical cancellation. The resulting discrete system is solved using a marching-on-in-time scheme in conjunction with iterative solvers. Numerical experiments for simply- and multiply-connected dielectric scatterers, including highly non-smooth geometries, corroborate the stability and efficiency of the proposed approach and demonstrate its ability to produce accurate derived quantities in the large-timestep regime.
\end{abstract}



\section{Introduction}
\IEEEPARstart{T}{he} numerical computation of transient electromagnetic fields scattered by piecewise homogeneous dielectric bodies often relies on solving time-domain boundary integral equations. Among several formulations for dielectric scattering, the time-domain Poggio--Miller--Chang--Harrington--Wu--Tsai (TD-PMCHWT) equation \cite{PM1973} and the time-domain (TD-) M\"{u}ller equation \cite{Muller1969} are most widely used. The TD-PMCHWT equation, like the time-domain electric field integral equation (TD-EFIE) for perfect electric conductors, is a first-kind integral equation. In contrast, the TD-M\"{u}ller equation, similar to the time-domain magnetic field integral equation (TD-MFIE), is of the second kind. Despite its ill-conditioned nature, the TD-PMCHWT equation is more widely used in practice, as it generally yields more accurate numerical solutions than the TD-M\"{u}ller formulation (see \cite{YTJ2008,YJN2013} for supporting arguments in the frequency-domain). In this work, we focus on developing a numerical method for the TD-PMCHWT equation that addresses three numerical challenges: ill-conditioning, late-time instability, and loss of accuracy in the derived quantities in the large-timestep regime. Here, the large-timestep regime refers to scenarios where relatively large timesteps are used in the time-domain simulation of the electromagnetic system. This regime resembles the low-frequency regime in the frequency-domain, since both are affected by similar numerical issues.

Due to the unbounded spectrum of the TD-EFIE operators appearing in the diagonal blocks, the TD-PMCHWT operator produces ill-conditioned matrix systems upon discretization, particularly for dense spatial meshes or large timesteps. These issues are commonly referred to as dense-mesh and large-timestep breakdowns, respectively. In the dense-mesh regime, the accumulation of eigenvalues near zero and infinity causes the ill-conditioning. This breakdown can be effectively resolved through operator preconditioning \cite{Hiptmair2006}, with Calder\'{o}n-based techniques being especially prominent. Several such methods have been developed for the frequency-domain (FD-) PMCHWT equation \cite{YJN2010a,CAM2011,NN2012,KBH+2022}, relying on either the PMCHWT operator itself or the electric field integral operator (EFIO) with a complex wavenumber. In the time-domain, frequency-domain operators with zero or purely imaginary wavenumbers can be used to build preconditioners, avoiding complex-valued systems and suppressing interior resonances \cite{LCA+2023,LCA+2024}. Alternatively, the TD-PMCHWT operator itself, when appropriately discretized, can serve as a dense-mesh preconditioner \cite{CAO+2009b}. However, using such a time-dependent operator requires an additional temporal discretization, leading to a more complicated marching-on-in-time (MOT) scheme.

Large-timestep breakdown, similar to low-frequency breakdown, stems from the opposite scaling behavior of matrix blocks associated with the loop and star components. While Calder\'{o}n preconditioners can also be employed to address this issue in the frequency-domain (by exploiting the anti-commutative property of boundary integral operators associated with the same medium \cite{YJN2010a}), this strategy becomes less effective in the time-domain setting. In particular, the necessary Calder\'{o}n anti-commutative relation does not hold between frequency-domain and time-domain operators, thus requiring nontrivial adaptations or dedicated implementation strategies. An alternative remedy is to rescale the Helmholtz blocks within the matrix system appropriately. This class of preconditioning has been extensively studied in the literature, both for frequency-domain formulations \cite{ACB+2013,BMC+2017,GAM+2017,MBC+2020,GSM+2025} and for their time-domain counterparts \cite{BCA2015,BCA2015b,BCA2015d,DAC2020,LCA+2024}.

In this work, we construct a diagonal dense-mesh preconditioner based on a modified static EFIO. The large-timestep preconditioner, on the other hand, is derived from the loop-star decomposition via the quasi-Helmholtz projectors combined with an appropriate rescaling scheme. Constructing the dense-mesh and large-timestep preconditioners separately offers a key advantage: each preconditioner is tailored to address its specific breakdown regime while remaining stable and well-conditioned in the presence of the other. Consequently, their combination yields a simple yet effective strategy for mitigating both dense-mesh and large-timestep breakdowns.

In addition to ill-conditioning, the TD-PMCHWT equation is also notorious for its late-time instability, which manifests as non-decaying and often exponentially growing errors at late times \cite{MB1982}. This phenomenon superficially resembles the direct-current instability of the TD-EFIE and originates from the constant-in-time solenoidal nullspace of the TD-EFIO. However, the presence of the double-layer operators in the off-diagonal blocks exacerbates the instability, making it highly sensitive to numerical quadrature errors \cite{LGC+2025}. Although various techniques can be used to suppress such errors \cite{SA1993,WPC+2004,SLY+2009,YE2006,VVV+2013,TXX2014}, these remedies generally do not guarantee stability, as they do not eliminate the underlying source of instability. In this work, we eliminate the source of instability by adopting the rescaling scheme introduced in \cite{BCA2015d}, which applies temporal differentiation and integration as rescaling factors associated with the Helmholtz components. This approach has been shown to effectively stabilize time-domain solutions for both the TD-EFIE \cite{BCA2015,BCA2015b} and TD-PMCHWT \cite{BCA2015d,LAC2024}, even under limited numerical precision in quadrature evaluations.

Lastly, in the low-frequency or large-timestep regime, the PMCHWT solution exhibits an imbalance in the scaling of its loop and star components. When these components are stored within the same floating-point variable, the dominated component experiences a loss of significant digits due to numerical cancellation against the dominant one \cite{ZC2000,CCS+2001,BCA+2014,GAM+2017}. Even though the field-trace solution may at first glance seem reasonable, large errors appear when computing secondary quantities of interest, such as the far field. The loss of numerical accuracy in the derived quantities in the large-timestep regime can also be mitigated through the loop-star decomposition combined with a rescaling scheme \cite{GAM+2017,BMC+2017}. In particular, when constructing the rescaling operators, we introduce a balancing factor for the dominated component, thereby defining a well-scaled auxiliary unknown that reduces the cancellation effects present in the original formulation.

In summary, this work addresses several numerical issues of the TD-PMCHWT equation, including dense-mesh breakdown, large-timestep breakdown, late-time instability, and loss of accuracy in the derived quantities in the large-timestep regime due to numerical cancellation. To combat dense-mesh breakdown, we introduce a multiplicative preconditioner utilizing a modified static EFIO. The remaining issues are simultaneously resolved by leveraging the loop-star decomposition combined with a rescaling scheme. The loop-star decomposition is performed via the quasi-Helmholtz projectors, which can be constructed without the need for costly identification of global loops \cite{Andriulli2012}. The rescaling operators are applied to the loop and star components, with temporal differentiation and integration used as rescaling factors. The resulting preconditioned and regularized formulation is temporally discretized using suitable pairs of trial and test functions. The associated matrix system is efficiently solved using the MOT algorithm in combination with iterative solvers, yielding stable time-domain solutions across all timestep sizes and significantly improving the accuracy of secondary quantity evaluation in the large-timestep regime. This paper extends our earlier conference contributions \cite{BCA2015,LAC2024} and the work in \cite{Beghein2015} by providing a rigorous analysis and a comprehensive suite of numerical results.

The structure of the paper is as follows. Section~\ref{sec:PMCHWT} gives an introduction to the TD-PMCHWT equation and its spatial Galerkin discretization. In Section~\ref{sec:preconditioning}, we develop a dense-mesh preconditioner for the TD-PMCHWT formulation. Section~\ref{sec:regulization} is devoted to asymptotic large-timestep analysis and a regularization strategy for the preconditioned system. Section~\ref{sec:discretization} discusses the temporal discretization and relevant implementation details, together with an analysis of the asymptotic computational complexity. Numerical results for both simply-connected and multiply-connected dielectric bodies, including a highly non-smooth domain, are presented in Section~\ref{sec:results}, corroborating the effectiveness of the proposed formulation. Finally, concluding remarks and future perspectives are provided in Section~\ref{sec:conclusion}.

\section{Time-Domain PMCHWT Equation}
\label{sec:PMCHWT}

Let $\Gm$ be the surface of an object in $\R^3$ which is filled by a homogeneous dielectric material with permittivity $\eps^\prime$ and permeability $\mu^\prime$. This object is immersed in a homogeneous background medium with permittivity $\epsilon$ and permeability $\mu$. The surface $\Gm$ is assumed to be closed, connected, and orientable, with the outward unit normal $\nv$. We consider the electromagnetic scattering of an incident transient plane wave $(\eb^{in}(\xb, t), \hb^{in}(\xb, t))$ at the boundary $\Gm$. The induced electric and magnetic surface current densities $\jb(\xb, t)$ and $\mb(\xb, t)$ are the solution to the following TD-PMCHWT equation\footnote{Time-domain continuous functions and operators are designated by calligraphic and normal symbols (e.g., $\TT, \KK, \jb, \mb, \eb, \hb$), while their semi-discrete and fully discrete counterparts are denoted by Fraktur (e.g., $\Tfrak, \Kfrak, \jfrakbm, \mfrakbm, \efrakbm, \hfrakbm$) and upright Roman (e.g., $\TTs, \KKs, \jjs, \ms, \ees, \hhs$) letters, respectively.}:
\begin{equation}
\label{eq:PMCHWT}
    \begin{pmatrix}
        \eta \TT + \eta^\prime \TT^\prime  & -\KK - \KK^\prime \\
        \KK + \KK^\prime & \frac{1}{\eta} \TT + \frac{1}{\eta^\prime} \TT^\prime
    \end{pmatrix}
    \begin{pmatrix}
        \jb \\
        \mb
    \end{pmatrix} 
    =
    \begin{pmatrix}
        \eb^{in} \times \nv \\
        \hb^{in} \times \nv
    \end{pmatrix},
\end{equation}
where the time-domain boundary integral operators $\TT$ and $\KK$ associated with the exterior region $(\eps, \mu)$ are defined by
\begin{align*}
    (\TT \jb)(\xb, t) & = (\TT^s \jb)(\xb, t) + (\TT^h \jb)(\xb, t), \\  
    (\TT^s \jb)(\xb, t) & = - \dfrac{1}{c} \, \nv \times \int_{\Gm} \dfrac{\pa_t \jb(\yb, \tau)}{4\pi \abs{\xb - \yb}} \ds_{\yb}, \\
    (\TT^h \jb)(\xb, t) & = c \, \nv \times \gradt_{\xxs} \int_{\Gm} \int_{-\infty}^{\tau}\dfrac{\divt_\Gm \, \jb(\yb, \xi)}{4\pi \abs{\xb - \yb}} \di \xi \ds_{\yb}, \\
    (\KK\jb)(\xb, t) & = \nv \times p.v. \int_{\Gm} \curlt_{\xxs} \dfrac{\jb(\yb, \tau)}{4\pi \abs{\xb - \yb}} \ds_{\yb}.
\end{align*}
Here, $\eta = \sqrt{\mu/\epsilon}, c = 1/\sqrt{\mu\epsilon}$, and $\tau = t - \abs{\xb - \yb}/c$. The operators and quantities $(\TT^\prime, \KK^\prime, \eta^\prime, c^\prime)$ associated with the interior region $(\eps^\prime, \mu^\prime)$ are defined analogously. We note that the causality is imposed to ensure the uniqueness of a solution to \eqref{eq:PMCHWT}, i.e., $\jb(\xb, t) = \zrb$ and $\mb(\xb, t) = \zrb$ for all $t < 0$ and $\xb$ in a neighborhood of $\Gm$.

The TD-PMCHWT equation \eqref{eq:PMCHWT} is first discretized in space. Let the boundary $\Gm$ be (approximately) triangulated into a mesh of $N_f$ flat triangles, with $N_e$ edges and $N_v$ vertices. On this mesh, we define a set of Rao--Wilton--Glisson (RWG) basis functions $\fb_n(\xb)$, with $n = 1, 2, \ldots, N_e$. Each RWG basis function is supported on a pair of adjacent triangles  $c^+_n$ and $c^-_n$ sharing a common edge $e_n$, and is defined by \cite{RWG1982}
\[
    \fb_n(\xb) = 
    \begin{cases}
        \dfrac{\xb - \xb^+_n}{2A_{c^+_n}} \qq & \text{for } \xb \in c^+_n, \\
        \dfrac{\xb^-_n - \xb}{2A_{c^-_n}} & \text{for } \xb \in c^-_n, \\
        \zrb & \text{otherwise},
    \end{cases}
\]
where $A_{c^+_n}$ and $A_{c^-_n}$ denote the areas of $c^+_n$ and $c^-_n$, respectively, and $\xb^+_n$ and $\xb^-_n$ are the vertices of $c^+_n$ and $c^-_n$ opposite to the common edge $e_n$ (see Fig.~\ref{fig:RWG_function}). The space spanned by the set of RWG basis functions $\brac{\fb_n}$ represents a div-conforming boundary element subspace of the corresponding spatial energy space 
\[
    \HHs^{-1/2}_{\times}(\divt_\Gm, \Gm) := \brac{\ub \in \HHs^{-1/2}_{\times}(\Gm) : \divt_{\Gm} \ub \in \Hs^{-1/2}(\Gm)}.
\]

\begin{figure}
    \centering
    \begin{tikzpicture}
      \draw[line width=0.4mm] (0, 0) -- (1, 1.73205) -- (2, 0) -- (1, -1.73205) -- (0, 0);
      \draw[line width=0.4mm, fill=gray!40] (2, 0) -- (3, 1.73205) -- (4, 0) -- (3, -1.73205) -- (2, 0);
      \draw[line width=0.4mm] (4, 0) -- (5, 1.73205) -- (6, 0) -- (5, -1.73205) -- (4, 0);
      \draw[line width=0.4mm] (1, 1.73205) -- (5, 1.73205);
      \draw[line width=0.4mm] (1, -1.73205) -- (5, -1.73205);
      \draw[line width=0.4mm] (0, 0) -- (6, 0);
      \draw[-latex, line width=0.6mm] (3, -0.6) -- (3, 0.6);
      \draw[fill=black] (3,1.73205) circle (2pt);
      \draw[fill=black] (3,-1.73205) circle (2pt);
      \draw[fill=black] (2,0) circle (2pt);
      \draw[fill=black] (4,0) circle (2pt);
      \node at (3.5, 0.2) {$e_n$};
      \node at (3.5, 1.55) {$\xb_n^-$};
      \node at (3.5, -1.48) {$\xb_n^+$};
      \node at (3, 1) {$c_n^-$};
      \node at (3, -0.95) {$c_n^+$};
    \end{tikzpicture}
    \caption{Rao--Wilton--Glisson basis function $\fb_n(\xb)$ associated with the edge $e_n$. The function $\fb_n(\xb)$ is supported on two adjacent triangles $c^-_n$ and $c^+_n$ sharing the common edge $e_n$. The bold arrow indicates the orientation of $\fb_n(\xb)$ across $e_n$, corresponding to a surface current flowing from $c^+_n$ to $c^-_n$.}
    \label{fig:RWG_function}
\end{figure}
The unknowns $\jb$ and $\mb$ are expanded in the RWG space as follows:
\[
    \jb(\xb, t) \approx \sum_{n=1}^{N_e} \jfrakbm_n(t) \fb_n(\xb), \q \mb(\xb, t) \approx \sum_{n=1}^{N_e} \mfrakbm_n(t) \fb_n(\xb).
\]
Their time-dependent coefficient vectors are denoted by
\begin{align*}
    \jfrakbm(t) & = \left(\jfrakbm_1(t), \jfrakbm_2(t), \ldots, \jfrakbm_{N_e}(t) \right)^\transpose, \\ 
    \mfrakbm(t) & = \left(\mfrakbm_1(t), \mfrakbm_2(t), \ldots, \mfrakbm_{N_e}(t) \right)^\transpose.
\end{align*}
Next, the equation \eqref{eq:PMCHWT} is tested with the rotated RWG functions $\nv \times \fb_m(\xb), m = 1, 2, \ldots, N_e$, giving rise to the following matrix system:
\begin{equation}
\label{eq:dis_PMCHWT}
    \begin{pmatrix}
        \Qfrak_{11} & \Qfrak_{12} \\
        \Qfrak_{21} & \Qfrak_{22}
    \end{pmatrix}
    \begin{pmatrix}
        \jfrakbm(t) \\
        \mfrakbm(t)
    \end{pmatrix} 
    =
    \begin{pmatrix}
        \efrakbm(t) \\
        \hfrakbm(t)
    \end{pmatrix},
\end{equation}
where the block matrix is defined by
\[
    \Qfrak :=
    \begin{pmatrix}
        \Qfrak_{11} & \Qfrak_{12} \\
        \Qfrak_{21} & \Qfrak_{22}
    \end{pmatrix}
    =
    \begin{pmatrix}
        \eta \Tfrak + \eta^\prime \Tfrak^\prime  & -\Kfrak - \Kfrak^\prime \\
        \Kfrak + \Kfrak^\prime & \frac{1}{\eta} \Tfrak + \frac{1}{\eta^\prime} \Tfrak^\prime
    \end{pmatrix},
\]
with the semi-discrete operators
\begin{align*}
    \Tfrak & = \Tfrak^s + \Tfrak^h, \\
    \left[(\Tfrak^s \jfrakbm)(t) \right]_m & = \sum\limits_{n=1}^{N_e} \inprod{\nv \times \fb_m, \TT^s \paren{\jfrakbm_n(t) \fb_n}}, \\
    \left[(\Tfrak^h \jfrakbm)(t) \right]_m & = \sum\limits_{n=1}^{N_e} \inprod{\nv \times \fb_m, \TT^h \paren{\jfrakbm_n(t) \fb_n}}, \\
    \left[(\Kfrak \, \jfrakbm)(t) \right]_m & = \sum\limits_{n=1}^{N_e} \inprod{\nv \times \fb_m, \KK \paren{\jfrakbm_n(t) \fb_n}},
\end{align*}
and the $\LLs^2(\Gm)$-inner product
\[
    \inprod{\fb, \gb} = \int_{\Gm} \fb(\xb) \cdot \gb(\xb) \ds.
\] 
The operators $\Tfrak^\prime$ and $\Kfrak^\prime$ are defined analogously. The right-hand side vectors are given by
\begin{align*}
    \left[\efrakbm(t)\right]_m & = \inprod{\nv \times \fb_m, \eb^{in} \times \nv}, \\ 
    \left[\hfrakbm(t)\right]_m & = \inprod{\nv \times \fb_m, \hb^{in} \times \nv}.
\end{align*}
Please note that the semi-discrete spatial discretization \eqref{eq:dis_PMCHWT} is not typically required in practical implementations. It is introduced here solely to motivate the time-dependent regularization proposed in Section~\ref{sec:regulization}. A fully space-time discretization of the TD-PMCHWT equation is presented in Section~\ref{sec:discretization}. Owing to the unboundedness of the PMCHWT operator spectrum, the semi-discrete system \eqref{eq:dis_PMCHWT} produces increasingly ill-conditioned matrices as the mesh density increases \cite{CAM2011}. In Section~\ref{sec:preconditioning}, we propose a preconditioner to alleviate this dense-mesh breakdown.

\section{Dense-Mesh Preconditioning}
\label{sec:preconditioning}

In this section, we construct a Calder\'{o}n dense-mesh preconditioner to the system \eqref{eq:dis_PMCHWT}, based on the following ``static EFIO'':
\begin{equation}
\label{eq:static_EFIE}
    T_0 = T^s_0 + T^h_0,
\end{equation}
where the singular and hyper-singular parts are defined by
\begin{align*}
    (T^s_0 \jb)(\xb) & = \dfrac{1}{D} \, \nv \times \int_{\Gm} \dfrac{\jb(\yb)}{4\pi \abs{\xb - \yb}} \ds_{\yb}, \\
    (T^h_0 \jb)(\xb) & = -D \,  \nv \times \gradt_{\xxs} \int_{\Gm} \dfrac{\divt_\Gm \, \jb(\yb)}{4\pi \abs{\xb - \yb}} \ds_{\yb}.
\end{align*}
Here, $D$ is the scatterer diameter (of length unit), included to ensure dimensional consistency between $T^s_0$ and $T^h_0$. In contrast to the standard frequency-domain EFIO, the wavenumber in the exponential kernels is set to zero, and the sign of $T_0^h$ is reversed. As a consequence, the operator $T_0$ is symmetric and elliptic on the energy trace space $\HHs^{-1/2}_{\times}(\divt_\Gm, \Gm)$ \cite{HS2003b}. When discretized using a Galerkin scheme, this property leads to symmetric positive-definite matrices.

To act as a dense-mesh preconditioner for \eqref{eq:dis_PMCHWT}, $T_0$ must be discretized in a dual space, in accordance with the operator preconditioning framework in \cite{Hiptmair2006}. Specifically, the discrete dual space must not only be a subspace of the energy trace space, but must also ensure that the Gram matrix coupling the primal RWG space with the dual space is uniformly inf-sup stable. A popular choice is the boundary element space spanned by the Buffa--Christiansen (BC) basis functions $\gb_n(\xb), n = 1, 2, \ldots, N_e$. These functions are defined as specific linear combinations of RWG basis functions supported on the barycentric refinement of the primal mesh (see Fig.~\ref{fig:BC_function} and \cite{BC2007}). Choosing the BC space as the discrete dual of the RWG space yields the following sparse, well-conditioned Gram matrix, which can be inverted efficiently using either a Krylov iterative solver or a direct solver:
\[
    \left[\GGs\right]_{mn} = \inprod{\nv \times \fb_m, \gb_n}.
\]

\begin{figure}
    \centering
    \begin{tikzpicture}
      \draw[line width=0.05mm, gray!30, fill=gray!45] (1, 0.57735) -- (1, -0.57735) -- (2, -1.1547) -- (3, -0.57735) -- (4, -1.1547) -- (5, -0.57735) -- (5, 0.57735) -- (4, 1.1547) -- (3, 0.57735) -- (2, 1.1547) -- (1, 0.57735) -- (1, 0.57735);
      \draw[line width=0.05mm, gray!30] (1, 1.73205) -- (1, -1.73205);
      \draw[line width=0.05mm, gray!30] (2, 1.73205) -- (2, -1.73205);
      \draw[line width=0.05mm, gray!30] (3, 1.73205) -- (3, -1.73205);
      \draw[line width=0.05mm, gray!30] (4, 1.73205) -- (4, -1.73205);
      \draw[line width=0.05mm, gray!30] (5, 1.73205) -- (5, -1.73205);
      \draw[line width=0.05mm, gray!30] (0, 0) -- (3, 1.73205);
      \draw[line width=0.05mm, gray!30] (6, 0) -- (3, 1.73205);
      \draw[line width=0.05mm, gray!30] (0, 0) -- (3, -1.73205);
      \draw[line width=0.05mm, gray!30] (6, 0) -- (3, -1.73205);
      \draw[line width=0.05mm, gray!30] (0.5, 0.866025) -- (5, -1.73205);
      \draw[line width=0.05mm, gray!30] (0.5, -0.866025) -- (5, 1.73205);
      \draw[line width=0.05mm, gray!30] (5.5, -0.866025) -- (1, 1.73205);
      \draw[line width=0.05mm, gray!30] (5.5, 0.866025) -- (1, -1.73205);
      \draw[line width=0.4mm] (0, 0) -- (1, 1.73205) -- (2, 0) -- (1, -1.73205) -- (0, 0);
      \draw[line width=0.4mm] (2, 0) -- (3, 1.73205) -- (4, 0) -- (3, -1.73205) -- (2, 0);
      \draw[line width=0.4mm] (4, 0) -- (5, 1.73205) -- (6, 0) -- (5, -1.73205) -- (4, 0);
      \draw[line width=0.4mm] (1, 1.73205) -- (5, 1.73205);
      \draw[line width=0.4mm] (1, -1.73205) -- (5, -1.73205);
      \draw[line width=0.4mm] (0, 0) -- (6, 0);
      \draw[-latex, line width=0.6mm] (2.35, 0) -- (3.65, 0);
      \node at (3, 0.2) {$e_n$};
    \end{tikzpicture}
    \caption{Buffa--Christiansen basis function $\gb_n(\xb)$ associated with the edge $e_n$. The function $\gb_n(\xb)$ is a linear combination of RWG basis functions supported on the barycentric refinement of the original mesh. The gray area denotes the support of $\gb_n(\xb)$, while the bold arrow indicates the direction of $\gb_n(\xb)$ along $e_n$.}
    \label{fig:BC_function}
\end{figure}

The operator $T_0$ is discretized using the BC functions into
\[
    \T_0 = \T_0^s + \T_0^h,
\]
with
\[
    \left[\T_0^s\right]_{mn} = \inprod{\nv \times \gb_m, T^s_0 \, \gb_n}, \q
    \left[\T_0^h\right]_{mn} = \inprod{\nv \times \gb_m, T^h_0 \, \gb_n}.
\]
The matrix $\T_0$ can be directly employed to construct a block-diagonal preconditioner for \eqref{eq:dis_PMCHWT}, with both diagonal blocks set to $\T_0$. The effectiveness of this preconditioner can be intuitively explained in light of the Calder\'{o}n identities. It is well known that the EFIO is self-preconditioning: when composed with itself, it yields a compact perturbation of the identity on $\HHs^{-1/2}_{\times}(\divt_\Gm, \Gm)$, which corresponds to the diagonal blocks of \eqref{eq:dis_PMCHWT} after preconditioning. Moreover, the double-layer operators appearing in the off-diagonal blocks of \eqref{eq:dis_PMCHWT} are compact on sufficiently smooth surfaces (in the frequency-domain). Because $\T_0$ is bounded on $\HHs^{-1/2}_{\times}(\divt_\Gm, \Gm)$, their composition with $\T_0$ preserves compactness, therefore does not affect the spectral clustering induced by the diagonal preconditioning.
From a rigorous standpoint, even though the classical Calder\'{o}n identities do not hold for operators associated with different material parameters and the double-layer operators may lose compactness on non-smooth surfaces, the preconditioning effect of $\T_0$ can still be justified within the general operator-preconditioning framework developed in \cite{Hiptmair2006}. In particular, any operator that is bounded and coercive on the underlying energy space, and that is discretized using a compatible dual space ensuring uniformly inf-sup stable Gram matrix, can serve as an effective preconditioner for \eqref{eq:dis_PMCHWT}.

In this work, we instead develop a loop-star diagonal version of $\T_0$ that preserves its preconditioning properties and is additionally beneficial for the large-timestep regularization introduced in Section~\ref{sec:regulization}. To this end, we recall the Helmholtz decomposition of the energy space, also known as the Hodge decomposition for boundary trace functions
\[
    \HHs^{-1/2}_{\times}(\divt_\Gm, \Gm) = \XXs_\Gm \oplus \HHs^{-1/2}_{\times}(\divt_\Gm0, \Gm),
\]
where $\XXs_\Gm$ and $\HHs^{-1/2}_{\times}(\divt_\Gm0, \Gm)$ denote the subspaces of irrotational (star) and solenoidal (loop) functions, respectively \cite{Buffa2001,BC2001b}. In the discrete setting using the BC basis, the quasi-Helmholtz projectors $\Pbb^\Lambda$ and $\Pbb^{\Sigma H}$ provide an approximation of the continuous Helmholtz decomposition. More specifically, these projectors map a function in the BC space to its quasi-irrotational and solenoidal parts, respectively (see \cite{ACB+2013} and also Section~\ref{sec:regulization} of this paper for their explicit definitions). Moreover, they are $l^2$-orthogonal, i.e., $\Pbb^{\Sigma H} \Pbb^\Lambda = \zrb$. 

Given that the operators $T^s_0$ and $T^h_0$ both define symmetric and elliptic bilinear forms when restricted to and tested with functions in $\HHs^{-1/2}_{\times}(\divt_\Gm0, \Gm)$ and $\XXs_\Gm$, respectively \cite{BHV+2003}, we employ the projectors $\Pbb^\Lambda$ and $\Pbb^{\Sigma H}$ to isolate the principal components of $\T_0$, resulting in a block-diagonal matrix with respect to the loop-star decomposition. In particular, we define
\begin{equation}
\label{eq:CP}
    \T = \T^{\Sigma H} + \T^\Lambda := \Pbb^{\Sigma H} \T_0^s \Pbb^{\Sigma H} + \Pbb^\Lambda \T_0^h \Pbb^\Lambda.
\end{equation}
The matrix $\T$ is symmetric and positive-definite, like $\T_0$. Consequently, $\T$ can serve as a dense-mesh preconditioner for the semi-discrete system \eqref{eq:dis_PMCHWT}. The resulting dense-mesh preconditioned formulation then reads
\begin{equation}
\label{eq:CP_PMCHWT}
    \T \GGs^{-1}
    \begin{pmatrix}
        \Qfrak_{11} & \Qfrak_{12} \\
        \Qfrak_{21} & \Qfrak_{22}
    \end{pmatrix}
    \begin{pmatrix}
        \jfrakbm(t) \\
        \mfrakbm(t)
    \end{pmatrix} 
    =
    \T \GGs^{-1}
    \begin{pmatrix}
        \efrakbm(t) \\
        \hfrakbm(t)
    \end{pmatrix}.
\end{equation}
Here and throughout the paper, the multiplication of a block matrix (e.g., $\Qfrak$) with block-sized matrices (e.g., $\T$ and $\GGs^{-1}$) is understood in a block-wise sense. The system \eqref{eq:CP_PMCHWT}, when accompanied by $\GGs^{-\transpose}$, produces well-conditioned matrices (in terms of the spectral condition number), independently of the mesh refinement and of the basis chosen for the finite element space \cite{Hiptmair2006}. The preconditioner $\T$ is not only effective in bounding the spectral condition number with respect to mesh density, but also immune to other numerical issues, including  large-timestep breakdown, numerical cancellation at large timesteps, and resonant instability. 

We remark that the static operator $T_0$ may be replaced by any EFIO $T_{-\iota\kappa}$ with an imaginary wavenumber $-\iota\kappa, \kappa > 0$ \cite{LC2024}. It is worth emphasizing that a bounded spectral condition number of the matrix on the left-hand side of \eqref{eq:CP_PMCHWT} does not guarantee boundedness of either its singular-value condition number or the number of iterations required by Krylov solvers to reach a prescribed tolerance. Nevertheless, numerical evidence indicates that \eqref{eq:CP_PMCHWT} exhibits a stable iteration count under mesh refinement, demonstrating its practical effectiveness.

In this sense, the formulation \eqref{eq:CP_PMCHWT} is well-conditioned with respect to spatial discretization. However, it remains time dependent and requires a temporal discretization. In addition, its time discretization still suffers from large-timestep breakdown, late-time instability, and numerical cancellation in the large-timestep regime, necessitating a regularization.

\section{Quasi-Helmholtz Regularization}
\label{sec:regulization}

This section presents an analysis on the asymptotic large-timestep scaling behavior of the dense-mesh preconditioned formulation \eqref{eq:CP_PMCHWT}. The goal is to investigate the mechanisms responsible for ill-conditioning and numerical cancellation at large timesteps. Based on these insights, we develop a regularization strategy that simultaneously mitigates large-timestep breakdown, suppresses late-time instability, and improves numerical accuracy of the derived quantities in the large-timestep regime.

To that end, we assume the loop-star decomposition is employed, resulting in the transformation matrices $\Lmdb, \HHs$, and $\Smb$. These matrices respectively map local loops, global loops (quasi-harmonic functions), and stars to discrete functions. Particularly, in the context of the RWG basis, the matrix $\Lmdb \in \R^{N_e \times N_v}$ is constructed via the node-edge adjacency information as follows:
\[
    \left[\Lmdb\right]_{mn} :=
    \begin{cases}
        1 \q & \text{if the vertex } n \text{ is identical with } v^+_m, \\
        -1 & \text{if the vertex } n \text{ is identical with } v^-_m, \\
        0 & \text{otherwise},
    \end{cases}
\]
where the edge $e_m$, $m = 1, 2, \ldots, N_e$, is oriented from the vertex $v_m^-$ to $v_m^+$. The matrix $\Smb \in \R^{N_e \times N_f}$ is linked to face-edge connectivity as follows:
\[
    \left[\Smb\right]_{mn} :=
    \begin{cases}
        1 \q & \text{if the triangle } n \text{ is identical with } c^+_m, \\
        -1 & \text{if the triangle } n \text{ is identical with } c^-_m, \\
        0 & \text{otherwise},
    \end{cases}
\]
where the RWG basis function $\fb_m(\xb), m = 1, 2, \ldots, N_e,$ represents a current flowing from the triangle $c_m^+$ to $c_m^-$ \cite{Andriulli2012}.

Both the local loops and the global loops represent solenoidal functions. When the surface $\Gm$ is simply-connected, the matrix $\HHs = \emptyset$. We emphasize that $\HHs$ is introduced solely for the analysis purpose. In our numerical scheme, the costly explicit identification of $\HHs$ is not required. Throughout this section, the large-timestep scaling of vector fields and matrices is with respect to the loop-star basis $(\Lmdb \,\, \HHs \,\, \Smb)^\transpose$.

\subsection{Large-Timestep Scaling}

The large-timestep scaling of physical quantities in time-domain is similar to their low-frequency counterparts in the frequency-domain setting. For consistency between time- and frequency-domain, we use the notation $\omega := \Delta t^{-1}$ in further large-timestep analysis, where $\Delta t$ denotes the timestep used in the simulation. In the large-timestep regime, the plane-wave excitation and its induced surface current densities scale as follows \cite{ZC2000,GAM+2017,BMC+2017}:
\begin{align*}
    \efrakbm \sim \hfrakbm & = \OO(\omega, \omega, 1)^\transpose, \\
    \jfrakbm \sim \mfrakbm & = \OO(1, \omega, \omega)^\transpose.
\end{align*}
When $\omega \to 0$ (equivalently, $\Delta t \to \infty$), the unknowns $\jfrakbm$ and $\mfrakbm$ are dominated by their local loop components. As a consequence, the global loop and star components suffer from a severe loss of significant digits when stored together with the local loops in the same floating-point variable, due to finite-precision arithmetic. This numerical cancellation leads to inaccuracies that may not be apparent in the solution itself or in the induced near fields, but becomes clearly visible in post-processed quantities, such as far-field patterns \cite{LCA+2024}.

For time-domain matrices, their large-timestep scaling can be derived from those of the TD-EFIE and TD-MFIE operators in \cite{LCA+2024}, or from the low-frequency counterpart in the frequency-domain \cite{GAM+2017,BMC+2017}. The TD-PMCHWT matrix $\Qfrak$ in \eqref{eq:dis_PMCHWT} has the following large-timestep scaling:
\[
    \begin{pmatrix}
        \Qfrak_{11} & \Qfrak_{12} \\
        \Qfrak_{21} & \Qfrak_{22}
    \end{pmatrix}
    =
    \OO\paren{
    \begin{array}{c c c | c c c}
        \omega & \omega & \omega & \omega^2 & \omega^2 & 1\\
        \omega & \omega & \omega & \omega^2 & 1 & 1 \\
        \omega & \omega & \omega^{-1} & 1 & 1 & 1 \\ \hline        
        \omega^2 & \omega^2 & 1 & \omega & \omega & \omega \\
        \omega^2 & 1 & 1 & \omega & \omega & \omega \\
        1 & 1 & 1 &\omega & \omega & \omega^{-1}
    \end{array}
    }.
\]
The opposite scaling behaviors $\OO(\omega)$ and $\OO(\omega^{-1})$ in the diagonal blocks, stemming from temporal differentiation and integration in the TD-EFIOs, are the origin of large-timestep breakdown. More specifically, the condition number of $\Qfrak$ grows as $\OO(\omega^{-1})/\OO(\omega) = \OO(\omega^{-2})$ when $\omega \to 0$.

Next, we derive the large-timestep scaling of the dense-mesh preconditioned formulation \eqref{eq:CP_PMCHWT}. Note that the dense-mesh preconditioner $\T$ and the Gram matrix $\GGs$ are independent of timestep $\Delta t$. In addition, in the BC dual basis (used for discretizing $\T$ and $\GGs$), the role of $\Lmdb$ and $\Smb$ are exchanged (i.e., $\Lmdb$ represents the stars and $\Smb$ represents the local loops), while $\HHs$ still represents the global loops \cite{ACB+2013}. Due to its loop-star diagonal structure, $\T$ behaves as follows in the basis $(\Lmdb \,\, \HHs \,\, \Smb)^\transpose$:
\[
    \T =
    \begin{pmatrix}
        \square & 0 & 0 \\
        0 & \square & \square \\
        0 & \square & \square
    \end{pmatrix},
\]
where $\square$ indicates a non-zero block. According to \cite{BCA2015b}, the inverse Gram matrix $\GGs^{-1}$ has the following structure:
\[
    \GGs^{-1} =
    \begin{pmatrix}
        \square & 0 & 0 \\
        \square & \square & 0 \\
        \square & \square & \square
    \end{pmatrix}.
\]
Combining the large-timestep scaling and algebraic structure of vector fields and matrices, the large-timestep scaling of the formulation \eqref{eq:CP_PMCHWT} reads as follows:
\begin{equation}
    \label{eq:scaling_DCP_PMCHWT}
    \paren{
    \begin{array}{c c c | c c c}
        \omega & \omega & \omega & \omega^2 & \omega^2 & 1\\
        \omega & \omega & \omega^{-1} & 1 & 1 & 1 \\
        \omega & \omega & \omega^{-1} & 1 & 1 & 1 \\ \hline 
        \omega^2 & \omega^2 & 1 & \omega & \omega & \omega \\
        1 & 1 & 1 & \omega & \omega & \omega^{-1} \\
        1 & 1 & 1 &\omega & \omega & \omega^{-1}
    \end{array}
    }
    \begin{pmatrix}
        1 \\
        \omega \\
        \omega \\ \hline
        1 \\
        \omega \\
        \omega \\
    \end{pmatrix}
    =
    \begin{pmatrix}
        \omega \\
        1 \\
        1 \\ \hline
        \omega \\
        1 \\
        1 \\
    \end{pmatrix}.
\end{equation}
It is therefore evident that \eqref{eq:CP_PMCHWT} still suffers from large-timestep breakdown due to the presence of the factor $\OO(\omega^{-1})$. In addition, numerical cancellation persists in the large-timestep regime. In Section~\ref{sec:lts_regularization}, we introduce a regularization for \eqref{eq:CP_PMCHWT} based on its large-timestep scaling behavior in \eqref{eq:scaling_DCP_PMCHWT}.

\subsection{Large-Timestep Regularization}
\label{sec:lts_regularization}

We recall the quasi-Helmholtz projectors associated with the RWG and BC bases
\begin{align*}
    \PPo^\Sm & := \Smb (\Smb^{\transpose} \Smb)^+ \Smb^\transpose, \q
    && \PPo^{\Lambda H} := I - \PPo^\Sm, \\
    \Pbb^\Lambda & := \Lmdb (\Lmdb^{\transpose} \Lmdb)^+ \Lmdb^\transpose, 
    && \, \Pbb^{\Sm H} := I - \Pbb^\Lambda,
\end{align*}
where the subscript $^+$ indicates the Moore--Penrose pseudo-inverse and $I$ is the identity matrix of size $N_e$ \cite{ACB+2013}. Roughly speaking, $\PPo^{\Lambda H}$ and $\PPo^\Sm$ act as projectors from the RWG space onto its loop and star subspaces, respectively. Similarly, $\Pbb^{\Sm H}$ and $\Pbb^\Lambda$ are the corresponding projectors associated with the BC space. Notably, this construction of the loop-star decomposition avoids the costly explicit identification of the global loops $\HHs$. A potential drawback is that local and global loops are not distinguished. However, as will be demonstrated in the subsequent analysis, this limitation is unproblematic.

Employing the quasi-Helmholtz projectors, we now propose a large-timestep regularization for \eqref{eq:CP_PMCHWT}. First, the star and loop components of the currents $\jfrakbm$ and $\mfrakbm$ are rescaled with respect to each other to balance their scalings in the large-timestep regime. To this end, we introduce the following right rescaling operator:
\begin{equation}
\label{eq:RSO}
    \RRs := \PPo^{\Lambda H} + \, \alpha \PPo^\Sm,
\end{equation}
where the dimensionless factor $\alpha = \OO(\om)$ is specified later. Obviously, $\RRs$ is invertible and its inverse is given by
\[
    \RRs^{-1} = \PPo^{\Lambda H} + \alpha^{-1} \PPo^\Sm.
\] 
The operator $\RRs$ is called the right rescaling operator because it is applied to the right of the TD-PMCHWT system matrix. The rescaling of the currents $\jfrakbm$ and $\mfrakbm$ is realized through the introduction of two auxiliary unknowns
\begin{equation}
\label{eq:xy}
    \ufrakbm(t) := \RRs^{-1} \jfrakbm(t), \qqq \vfrakbm(t) := \RRs^{-1} \mfrakbm(t),
\end{equation}
which scale as follows when $\om \to 0$:
\[
    \ufrakbm \sim \vfrakbm = \OO(1, \omega, 1)^{\transpose}.
\]
It is observed that, in $\ufrakbm$ and $\vfrakbm$, the star component $\Smb$ is balanced with the local loop component $\Lmdb$, while the global loop component $\HHs$ remains dominated. Although $\HHs$ has fixed rank, this imbalance still causes numerical cancellation. Nevertheless, this cancellation does not severely affect the computation of the scattered fields, since the contribution of $\HHs$ is only of order $\OO(\omega)$ relative to that of $\Lmdb$. Moreover, this behavior is physically consistent, as a plane wave does not excite global loop currents in the static limit \cite{BMC+2017}. Other types of excitation may lead to a complete balance among the three components of the rescaled unknowns. Further details on the low-frequency scaling behavior under different excitations can be found in \cite{HEA+2023}.

By introducing the rescaling operator $\RRs$, the original system in terms of the physical unknowns $\jfrakbm$ and $\mfrakbm$ is transformed into an equivalent system involving the auxiliary unknowns $\ufrakbm$ and $\vfrakbm$. This transformation significantly mitigates the imbalance among the solution components and thereby improves numerical accuracy in the large-timestep regime. The physical currents $\jfrakbm$ and $\mfrakbm$ can be recovered straightforwardly by applying $\RRs$ to $\ufrakbm$ and $\vfrakbm$, i.e.,
\begin{equation}
\label{eq:jm}
    \jfrakbm(t) = \RRs \ufrakbm(t), \qqqq
    \mfrakbm(t) = \RRs \vfrakbm(t).
\end{equation}

To eliminate the large-timestep ill-conditioning of the system matrix in \eqref{eq:CP_PMCHWT}, we additionally introduce the following left rescaling operator:
\begin{equation}
\label{eq:LSO}
    \Lbb := \Pbb^{\Sigma H} + \alpha^{-1} \Pbb^\Lambda.
\end{equation}
The introduction of the left rescaling operator can also be interpreted as a rescaling of the right-hand side. This operator is constructed from the quasi-Helmholtz projectors $\Pbb^{\Sigma H}$ and $\Pbb^\Lambda$ associated with the BC basis, in contrast to the right rescaling operator $\RRs$, which is defined on the RWG space. Now, applying $\Lbb$ to the left of \eqref{eq:CP_PMCHWT} and substituting the physical unknowns $\jfrakbm$ and $\mfrakbm$ using \eqref{eq:jm}, we obtain the following system in terms of the auxiliary unknowns $\ufrakbm$ and $\vfrakbm$:
\begin{equation}
    \label{eq:reg_DCP_PMCHWT}
    \begin{pmatrix}
        \wt{\Qfrak}_{11} & \wt{\Qfrak}_{12} \\
        \wt{\Qfrak}_{21} & \wt{\Qfrak}_{22}
    \end{pmatrix}
    \begin{pmatrix}
        \ufrakbm(t) \\
        \vfrakbm(t)
    \end{pmatrix} 
    =
    \begin{pmatrix}
        \wt{\efrakbm}(t) \\
        \wt{\hfrakbm}(t)
    \end{pmatrix},
\end{equation}
where the matrix blocks
\[ 
    \wt{\Qfrak}_{ij} = \Lbb \T \GGs^{-1} \Qfrak_{ij} \RRs,
\]
with $i, j = 1, 2$, and the right-hand side
\[
    \wt{\efrakbm} = \Lbb \T \GGs^{-1} \efrakbm, \qqq \wt{\hfrakbm} = \Lbb \T \GGs^{-1} \hfrakbm.
\]
Finally, straightforward scaling arguments yield the following large-timestep behavior of \eqref{eq:reg_DCP_PMCHWT}:
\[
    \left(
    \begin{array}{c c c | c c c}
        1 & 1 & \omega & \omega & \omega & 1\\
        \omega & \omega & 1 & 1 & 1 & \omega \\
        \omega & \omega & 1 & 1 & 1 & \omega \\ \hline
        \omega & \omega & 1 & 1 & 1 & \omega \\
        1 & 1 & \omega & \omega & \omega & 1 \\
        1 & 1 & \omega & \omega & \omega & 1
    \end{array}
    \right)
    \begin{pmatrix}
        1 \\
        \omega \\
        1 \\ \hline
        1 \\
        \omega \\
        1 \\
    \end{pmatrix}
    =
    \begin{pmatrix}
        1 \\
        1 \\
        1 \\ \hline
        1 \\
        1 \\
        1 \\
    \end{pmatrix}.
\]
Notably, no factor $\OO(\om^{-1})$ appears in the scaling of the system matrix in \eqref{eq:reg_DCP_PMCHWT}. Numerical experiments indicate that its condition number remains bounded as $\omega \to 0$. Consequently, the formulation \eqref{eq:reg_DCP_PMCHWT} is free from large-timestep breakdown, and its solution is well balanced in the large-timestep regime.

The left and right rescaling operators $\Lbb$ and $\RRs$ are essential for the construction of \eqref{eq:reg_DCP_PMCHWT}. These operators differ from those proposed for the FD-PMCHWT formulation in \cite{BMC+2017}, as symmetry between the two components is not required here. Nevertheless, their effectiveness remains the same as those in \cite{BMC+2017}, provided that $\alpha = \OO(\om)$. 

In the time-domain setting, there are two possible choices of $\alpha$ satisfying this criterion, either $\alpha = T_{\max} \Delta t^{-1}$ or $\alpha = T_{\max} \pa_t$, where the constant $T_{\max}$ of time unit is included to render $\alpha$ dimensionless \cite{LCA+2024}. The latter choice is motivated by the observation that, for a slowly varying function, the time derivative scales as $\OO(\Delta t^{-1})$, as follows from a Taylor expansion. If the focus is restricted to conditioning and numerical cancellation, which are direct counterparts of the challenges encountered in the frequency-domain, then either choice of $\alpha$ is sufficient. In Section~\ref{sec:stabilization}, however, time-domain specific problems will motivate a more careful selection.

It is further noted that the left rescaling operator $\Lbb$ and the dense-mesh preconditioner $\T$ commute due to the loop-star diagonal structure of $\T$, i.e.,
\[
    \Lbb \T = \T \Lbb = \Pbb^{\Sigma H} \T_0^s \Pbb^{\Sigma H} + \alpha^{-1} \Pbb^\Lambda \T_0^h \Pbb^\Lambda.
\]
As a result, their positions in \eqref{eq:reg_DCP_PMCHWT} can be interchanged, yielding a dense-mesh preconditioned (by $\T$) version of the regularized TD-PMCHWT formulation proposed in \cite{BCA2015d}. The preconditioned and regularized formulation \eqref{eq:reg_DCP_PMCHWT} also differs from that presented in \cite{Beghein2015}, where the regularized TD-PMCHWT system was preconditioned using its dual counterpart.

\subsection{Late-Time Stabilization}
\label{sec:stabilization}

Up to now, we have addressed three of the four numerical issues associated with the TD-PMCHWT formulation: dense-mesh breakdown, large-timestep breakdown, and numerical cancellation in the large-timestep regime. This section focuses on the remaining issue, namely late-time instability, which is known to originate from the nullspace of the TD-EFIOs consisting of constant-in-time solenoidal functions \cite{LGC+2025}.

Our remedy consists in getting rid of the nullspace of the TD-EFIOs by choosing an appropriate rescaling factor $\alpha$ in \eqref{eq:RSO} and \eqref{eq:LSO}. More specifically, we set
\begin{equation}
\label{eq:alpha}
    \alpha = \tilde{\pa}_t := T_{\max} \pa_t, \qq \alpha^{-1} = \tilde{\pa}_t^{-1} := \pa^{-1}_t/T_{\max},
\end{equation}
where $\pa_t^{-1}$ is the temporal integration over $(-\infty, t)$ and $T_{\max} = D/c_{\min}$, with $c_{\min} := \min(c, c^\prime)$. The idea of employing temporal differentiation and integration as the rescaling factors was originally introduced in \cite{BCA2015,BCA2015b} for the TD-EFIE. In the following, we demonstrate how this approach can be adapted to the TD-PMCHWT equation. It is important to note that the low-frequency and large-timestep regularization of the PMCHWT differs fundamentally from that of the EFIE. In particular, the left rescaling operator for the EFIE can be constructed from the projectors associated with the RWG space \cite{BCA2015}, whereas the corresponding operator for the PMCHWT is defined on the BC space. The necessity of employing a distinct regularization scheme for the PMCHWT equation has been explained in \cite{BMC+2017}.

With the choice of $\alpha$ in \eqref{eq:alpha}, we consider the diagonal blocks $\wt{\Qfrak}_{11}$ and $\wt{\Qfrak}_{22}$ in \eqref{eq:reg_DCP_PMCHWT} that contain the TD-EIFOs $\TT$ and $\TT^\prime$, with a particular attention to the contributions from $\TT$ (as $\TT^\prime$ can be handled analogously). We have the following decomposition:
\begin{equation}
\label{eq:decomposition}
    \Zfrak := \Lbb \T \GGs^{-1} \Tfrak \RRs = \Zfrak_{ll} + \Zfrak_{ls} + \Zfrak_{sl} + \Zfrak_{ss},
\end{equation}
where the components differ by the left and right projectors
\begin{align*}
    \Zfrak_{ll} & := \T^{\Sigma H} \GGs^{-1} \Tfrak^s \PPo^{\Lambda H}, \\
    \Zfrak_{ls} & := \T^{\Sigma H} \GGs^{-1} \tilde{\pa}_t \Tfrak \PPo^\Sm, \\
    \Zfrak_{sl} & := \T^\Lambda \GGs^{-1} \PPo^{\Lambda H} \tilde{\pa}^{-1}_t \Tfrak^s \PPo^{\Lambda H}, \\
    \Zfrak_{ss} & := \T^\Lambda \GGs^{-1} \PPo^{\Lambda H} \Tfrak^s \PPo^\Sm.
\end{align*}
Here, we have used the identities $\PPo^{\Lambda H} \Tfrak^h = \Tfrak^h \PPo^{\Lambda H} = \zrb$. The projector $\PPo^{\Lambda H}$ appears in the middle  of the last two terms because we have leveraged the identity $\Pbb^\Lambda \GGs^{-1} \PPo^\Sm = \zrb$ from \cite{BCA2015b}. One can see that the rescaling temporal operators $\tilde{\pa}_t$ and $\tilde{\pa}^{-1}_t$ appear only in the components $\Zfrak_{ls}$ and $\Zfrak_{sl}$. In particular, $\tilde{\pa}^{-1}_t$ cancels the temporal differentiation present in $\Tfrak^s$. As a consequence, the kernel of $\Zfrak_{sl}$ contains no non-zero constant-in-time solenoidal functions, and therefore neither does that of $\Zfrak$. In other words, $\Zfrak$ possesses only a trivial nullspace. The operator $\tilde{\pa}_t$ also cancels the temporal integration in $\Tfrak^h$, whose benefit will be clarified in Section~\ref{sec:computation}.

To conclude this section, the preconditioned and regularized formulation \eqref{eq:reg_DCP_PMCHWT}, with $\alpha$ chosen in \eqref{eq:alpha}, is immune to ill-conditioning, late-time instability, and numerical cancellation in the large-timestep regime. Since \eqref{eq:reg_DCP_PMCHWT} is still time-dependent, an appropriate temporal discretization scheme is needed.
 
\section{Discretization and Implementation}
\label{sec:discretization}

\subsection{Temporal Discretization}
This section presents a temporal discretization of the regularized TD-PMCHWT formulation \eqref{eq:reg_DCP_PMCHWT}. In introducing the auxiliary unknowns $\ufrakbm$ and $\vfrakbm$, we have added a temporal integration $\tilde{\pa}_t^{-1}$ to the star components $\PPo^\Sm$ of $\jfrakbm$ and $\mfrakbm$, while retained their loops $\PPo^{\Lambda H}$. To ensure the compatibility between the components of physical unknowns $\jfrakbm$ and $\mfrakbm$, we discretize the stars and loops of $\ufrakbm$ and $\vfrakbm$ using two different sets of temporal basis functions, such that the set for the stars $\PPo^\Sm$ is of one order higher than that for the loops $\PPo^{\Lambda H}$. For simplicity, in this paper we opt for the lowest-order pair.

Let $N_t$ be the number of time steps we want to compute for the numerical solution. The auxiliary unknowns $\ufrakbm$ and $\vfrakbm$ are approximated as follows:
\[
    \begin{pmatrix}
        \ufrakbm(t) \\
        \vfrakbm(t)
    \end{pmatrix} 
    =
    \sum\limits_{i = 1}^{N_t} \paren{p_i(t) \PPo^{\Lambda H} + \, h_i(t) \PPo^\Sm}
    \begin{pmatrix}
        \uv_i \\
        \vvs_i
    \end{pmatrix},
\]
where the two sets of temporal shifted basis functions
\begin{align*}
    & p_i(t) := p_0(t - i\Delta t), && \q p_0(t) = 
    \begin{cases}
        1 \,\, & -\Delta t < t < 0, \\
        0 & \text{otherwise},
    \end{cases} \\
    & h_i(t) := h_0(t - i\Delta t), && \q h_0(t) = 
    \begin{cases}
        1 - \dfrac{\abs{t}}{\Delta t} \,\, & \abs{t} \le \Delta t, \\
        0 & \text{otherwise},
    \end{cases}
\end{align*}
with $i = 1, 2, \ldots, N_t$ (see Fig.~\ref{fig:ph_function}). 

\begin{figure}[t]
    \centering
    \begin{tikzpicture}
        \draw[-latex,line width=0.2mm] (-6.5,-0.3)--(-3,-0.3) node[below=1mm]{$t$}; 
        \draw[-latex,line width=0.2mm] (-6.1,-0.6)--(-6.1,1.7) node[right]{$p_0(t)$};
        \draw (-5.2,-0.4)--(-5.2,-0.3) node[below=0.6mm]{$-\Delta t$};
        \draw (-4,-0.4)--(-4,-0.3) node[below=0.8mm]{$0$};
        \draw (-6.1,0)--(-6.2,0) node[left]{$0$};
        \draw (-6.1,1.2)--(-6.2,1.2) node[left]{$1$};
        \draw[line width=0.45mm,black] (-5.9,0) -- (-5.2,0);
        \draw[line width=0.45mm,black] (-5.2,1.2) -- (-4,1.2);
        \draw[line width=0.45mm,black] (-4,0) -- (-3.3,0);
        
        \draw[-latex,line width=0.2mm] (-2.5,-0.3)--(2.1,-0.3) node[below=1mm]{$t$}; 
        \draw[-latex,line width=0.2mm] (-2.1,-0.6)--(-2.1,1.7) node[right]{$h_0(t)$};
        \draw (-1.2,-0.4)--(-1.2,-0.3) node[below=0.6mm]{$-\Delta t$};
        \draw (0,-0.4)--(0,-0.3) node[below=0.8mm]{$0$};
        \draw (1.2,-0.4)--(1.2,-0.3) node[below=0.6mm]{$\Delta t$};
        \draw (-2.1,0)--(-2.2,0) node[left]{$0$};
        \draw (-2.1,1.2)--(-2.2,1.2) node[left]{$1$};
        \draw[line width=0.45mm,black] (-1.9,0) -- (-1.2,0) -- (0,1.2) -- (1.2,0) -- (1.9,0);
    \end{tikzpicture}
    \caption{Piecewise-constant basis function $p_0$ (left) and continuous piecewise-linear basis function $h_0$ (right).}
    \label{fig:ph_function}
\end{figure}

For the testing, a similar argument applies, with a note that the star component $\Pbb^\Lambda$ in \eqref{eq:LSO} contains an extra temporal integration due to rescaling. This, together with the compatibility
\begin{equation}
    \dfrac{1}{\Delta t} \int_\R p_i(t) \, p_0(t + \xi) \dt = \int_\R \delta_i(t) \, h_0(t + \xi) \dt, \label{eq:ph}
\end{equation}
which holds for all $\xi \in \R$ and $i = 0, 1, \ldots, N_t$, leads to the following test functions:
\[
    \dfrac{1}{\Delta t} \, p_k(t) \, \Pbb^{\Sigma H} + \delta_k(t) \,\Pbb^\Lambda.
\]
Here, $\delta_k(t) := \delta(t - k\Delta t), k = 1, 2, \ldots, N_t,$ are the Dirac delta distributions. Denoting $\wws_i := \paren{\uv_i, \vvs_i}^\transpose$, the temporal testing of \eqref{eq:reg_DCP_PMCHWT} yields a lower triangular block matrix system of the form
\begin{equation}
\label{eq:fulldis_PMCHWT}
    \begin{pmatrix}
        \QQs_0 & & & \\
        \QQs_1 & \QQs_0 & & \\
        \vdots & \vdots & \ddots & \\
        \QQs_{N_t-1} & \QQs_{N_t-2} & \ldots & \QQs_0
    \end{pmatrix}   
    \begin{pmatrix}
        \wws_1 \\
        \wws_2 \\
        \vdots \\
        \, \wws_{N_t}
    \end{pmatrix} 
    =
    \begin{pmatrix}
        \rrs_1 \\
        \rrs_2 \\
        \vdots \\
        \rrs_{N_t}
    \end{pmatrix}.
\end{equation}

Please note that this temporal discretization is highly non-standard, since the basis functions do not even possess a tensorial structure. In Sections~\ref{sec:computation} and \ref{sec:rhs}, we will compute the matrix blocks $\QQs_k$, with $k = 0, 1, \ldots, N_t-1,$ and the right-hand side vectors $\rrs_i, i = 1, 2, \ldots, N_t,$ using a more conventional scheme. Once the matrices and the right-hand sides are computed, the unknowns $\wws_i, i = 1, 2, \ldots, N_t,$ can be sequentially obtained by solving the MOT equation
\begin{equation}
    \label{eq:MOT}
    \QQs_0 \wws_i = \rrs_i - \sum\limits_{k=1}^{i-1} \QQs_k \wws_{i-k}.
\end{equation}
This equation can be efficiently solved by iterative solvers (e.g., GMRES), as the matrix $\QQs_0$ is well-conditioned when the spatial mesh size decreases or the timestep increases, as the result of preconditioning.

After solving the fully discrete system \eqref{eq:fulldis_PMCHWT} for the auxiliary unknowns, a post-processing procedure is performed to obtain the physical current densities. Following \cite{LCA+2024}, we have the following relation:
\begin{align*}
    \begin{pmatrix}
        \jfrakbm(t)  \\
        \mfrakbm(t)
    \end{pmatrix}
    & = \paren{\PPo^{\Lambda H} + \, \tilde{\pa}_t \PPo^\Sm} 
    \begin{pmatrix}
        \ufrakbm(t) \\
        \vfrakbm(t)
    \end{pmatrix} \\
    & = \sum\limits_{i=1}^{N_T} \paren{p_i(t) \PPo^{\Lambda H} + \, \tilde{\pa}_t h_i(t) \PPo^\Sm} 
    \begin{pmatrix}
        \uv_i \\
        \vvs_i
    \end{pmatrix} \\
    & = \sum\limits_{i=1}^{N_T} \paren{\PPo^{\Lambda H}     \begin{pmatrix}
        \uv_i \\
        \vvs_i
    \end{pmatrix} 
    + \dfrac{T_{\max}}{\Delta t} \PPo^\Sm  
    \begin{pmatrix}
        \uv_i - \uv_{i-1} \\
        \vvs_i - \vvs_{i-1}
    \end{pmatrix}} p_i(t),
\end{align*}
with the convention $\uv_0 = \vvs_0 = \zrb$. Let $\paren{\jjs_i, \ms_i}^\transpose$ be the expansion coefficient of $\paren{\jfrakbm(t), \mfrakbm(t)}^\transpose$ corresponding to the temporal basis function $p_i(t)$, with $i = 1, 2, \ldots, N_t$. Then, we can conclude
\begin{equation}
\label{eq:dis_jm}
    \begin{pmatrix}
        \jjs_i \\
        \ms_i
    \end{pmatrix}
    =
    \PPo^{\Lambda H} 
    \begin{pmatrix}
        \uv_i \\
        \vvs_i
    \end{pmatrix} 
    + \, \dfrac{T_{\max}}{\Delta t} \PPo^\Sm
    \begin{pmatrix}
        \uv_i -\uv_{i-1} \\
        \vvs_i - \vvs_{i-1}
    \end{pmatrix}.
\end{equation}
In fact, solution computed from \eqref{eq:dis_jm} still suffers from numerical cancellation at large timesteps $\Delta t$, as the star component $\PPo^\Sm$ experiences a loss of significant digits when stored together with the loop component $\PPo^{\Lambda H}$. In the far-field computation, the two components of $\jjs_i$ and $\ms_i$ should be stored separately to avoid numerical cancellation effects.

\subsection{Computation of Matrix Blocks}
\label{sec:computation}

In this section, we describe in detail how to compute the matrix blocks $\QQs_i$ in \eqref{eq:fulldis_PMCHWT} for practical implementation. First, we introduce the following continuously differentiable piecewise-quadratic function (see Fig.~\ref{fig:quad_function}):
\begin{align*}
    q_0(t) 
    & : = \dfrac{1}{\Delta t} \int_\R p_0(\xi) \, h_0(t + \xi) \di \xi \\
    & \, =
    \begin{cases}
        \dfrac{1}{2} \paren{\dfrac{t}{\Delta t} + 1}^2 & -\Delta t \le t < 0, \\[0.2cm]
        \dfrac{1}{2} - \dfrac{t^2}{\Delta t^2} + \dfrac{t}{\Delta t} &  0 \le t < \Delta t, \\[0.2cm]
        \dfrac{1}{2} \paren{\dfrac{t}{\Delta t} - 2}^2 \qq & \Delta t \le t \le 2\Delta t, \\
        0 & \text{otherwise}.
    \end{cases}
\end{align*}
\begin{figure}[t]
    \centering
    \begin{tikzpicture}
        \draw[-latex,line width=0.2mm] (-2.7,-0.3)--(3.3,-0.3) node[below=1mm]{$t$}; 
        \draw[-latex,line width=0.2mm] (-2.3,-0.6)--(-2.3,1.7) node[right]{$q_0(t)$};
        \draw (-1.2,-0.4)--(-1.2,-0.3) node[below=0.6mm]{$-\Delta t$};
        \draw (0,-0.4)--(0,-0.3) node[below=0.8mm]{$0$};
        \draw (1.2,-0.4)--(1.2,-0.3) node[below=0.6mm]{$\Delta t$};
        \draw (2.4,-0.4)--(2.4,-0.3) node[below=0.6mm]{$2\Delta t$};
        \draw (-2.3,0)--(-2.4,0) node[left]{$0$};
        \draw (-2.3,1.2)--(-2.4,1.2) node[left]{$1$};
        \draw[line width=0.45mm,black] (-1.9,0) -- (-1.2,0);
        \draw[domain=-1.2:0,smooth,variable=\x,line width = 0.45mm,black] plot ({\x},{0.6*(\x/1.2+1)^2});
        \draw[domain=0:1.2,smooth,variable=\x,line width = 0.45mm,black] plot ({\x},{0.6-\x^2/1.2+\x});
        \draw[domain=1.2:2.4,smooth,variable=\x,line width = 0.45mm,black] plot ({\x},{0.6*(\x/1.2-2)^2});
        \draw[line width=0.45mm,black] (2.4,0) -- (3,0);
    \end{tikzpicture}
    \caption{Continuously differentiable piecewise-quadratic basis function $q_0$.}
    \label{fig:quad_function}
\end{figure}
This function also satisfies the compatibility property
\begin{equation}
    \dfrac{1}{\Delta t} \int_\R p_i(t) \, h_0(t + \xi) \dt = \int_\R \delta_i(t) \, q_0(t + \xi) \dt.\label{eq:hq}
\end{equation}

The matrix blocks $\QQs_i$ in \eqref{eq:fulldis_PMCHWT} are defined by
\[
    \QQs_i = \int_\R \paren{\dfrac{p_i(t)}{\Delta t} \Pbb^{\Sigma H} + \delta_i(t) \Pbb^\Lambda} \wt{\Qfrak} \paren{p_0(t) \PPo^{\Lambda H} + h_0(t) \PPo^\Sm} \dt,
\]
with $i = 0, 1, \ldots, N_t-1,$ and $\wt{\Qfrak}$ the block matrix in \eqref{eq:reg_DCP_PMCHWT}, i.e.,
\[
    \wt{\Qfrak} = 
    \begin{pmatrix}
        \wt{\Qfrak}_{11} & \wt{\Qfrak}_{12} \\
        \wt{\Qfrak}_{21} & \wt{\Qfrak}_{22}
    \end{pmatrix}.
\]
Analogous to \eqref{eq:decomposition}, $\QQs_i$ can be decomposed into the following four matrices based on the left and right multipliers:
\begin{align*}     
    \QQs_i^{ll} & = \dfrac{1}{\Delta t} \Pbb^{\Sigma H} \int_\R  p_i(t) \wt{\Qfrak} \paren{p_0(t) \PPo^{\Lambda H}} \dt \\
                & = \, \Pbb^{\Sigma H} \int_\R \delta_i(t) \wt{\Qfrak} \paren{h_0(t) \PPo^{\Lambda H}} \dt, \\
    \QQs_i^{ls} & = \dfrac{1}{\Delta t} \Pbb^{\Sigma H} \int_\R  p_i(t) \wt{\Qfrak} \paren{h_0(t) \PPo^\Sm} \dt \\
                & = \, \Pbb^{\Sigma H} \int_\R \delta_i(t) \wt{\Qfrak} \paren{q_0(t) \PPo^\Sm} \dt, \\
    \QQs_i^{sl} & = \Pbb^\Lambda \int_\R \delta_i(t) \wt{\Qfrak} \paren{p_0(t) \PPo^{\Lambda H}} \dt, \\
    \QQs_i^{ss} & = \Pbb^\Lambda \int_\R \delta_i(t) \wt{\Qfrak} \paren{h_0(t) \PPo^\Sm} \dt.
\end{align*}
In the first two terms, we have used the properties \eqref{eq:ph} and \eqref{eq:hq}. Please note that each of the four components of $\QQs_i$ is a $2\times 2$ block matrix, inheriting the structure of $\wt{\Qfrak}$. 

Next, let us have a deeper look into each block of each matrix. In particular, we demonstrate how to compute $\QQs_i^{ll}, \QQs_i^{ls}, \QQs_i^{sl}$ and $\QQs_i^{ss}$ in practice. We note that the diagonal blocks $\wt{\Qfrak}_{11}$ and $\wt{\Qfrak}_{22}$ of $\wt{\Qfrak}$ only involve the TD-EFIOs $\TT$ and $\TT^\prime$, whereas the off-diagonal blocks $\wt{\Qfrak}_{12}$ and $\wt{\Qfrak}_{21}$ only involve $\KK$ and $\KK^\prime$. Without loss of generality, we focus on the contributions from $\TT$ and $\KK$ only. The corresponding counterparts of $\TT^\prime$ and $\KK^\prime$ are tagged with a superscript $^\prime$.

\subsubsection{The matrices $\QQs_i^{ll}$ and $\QQs_i^{ss}$}

The contributions from $\TT$ (to the diagonal blocks) of these matrices are in fact the temporal discretization of $\Zfrak_{ll}$ and $\Zfrak_{ss}$ in \eqref{eq:decomposition}, respectively, i.e.,
\begin{align*}
    \ZZs_i^{ll} & := \T^{\Sigma H} \GGs^{-1} \int_\R \delta_i(t) \Tfrak^s \paren{h_0(t) \PPo^{\Lambda H}} \dt, \\
    \ZZs_i^{ss} & := \T^\Lambda \GGs^{-1} \PPo^{\Lambda H} \int_\R \delta_i(t) \Tfrak^s \paren{h_0(t) \PPo^\Sm} \dt.
\end{align*}
In practice, it is unnecessary (and probably impossible) to first discretize $\TT^s$ in space to get $\Tfrak^s$, then discretize $\Tfrak^s$ in time. Instead, one can perform a collocation-in-time Galerkin-in-space discretization of $\TT^s$. Let $\TTs_i$ be the fully space-time discretization of $\TT^s$, whose elements are defined by
\begin{align*}
    \left[\TTs_i\right]_{mn} 
    & := \int_\R \delta_i(t) \inprod{\nv \times \fb_m, \TT^s\paren{h_0(t) \fb_n}} \dt \\
    & \ = \left.\inprod{\nv \times \fb_m, \TT^s\paren{h_0(t) \fb_n}}\right|_{t = i\Delta t},
\end{align*}
with $m, n =  1, 2, \ldots, N_e$. The matrices $\ZZs_i^{ll}$ and $\ZZs_i^{ss}$ can be rewritten as follows:
\begin{align*}
    \ZZs_i^{ll} & = \T^{\Sigma H} \GGs^{-1} \TTs_i \PPo^{\Lambda H}, \\
    \ZZs_i^{ss} & = \T^\Lambda \GGs^{-1} \PPo^{\Lambda H} \TTs_i \PPo^\Sm.
\end{align*}
Analogously, we define the matrix $\KKs_i$ such that
\begin{align*}
    \left[\KKs_i\right]_{mn} 
    & := \int_\R \delta_i(t) \inprod{\nv \times \fb_m, \KK \paren{h_0(t) \fb_n}} \dt \\
    & \ = \left.\inprod{\nv \times \fb_m, \KK \paren{h_0(t) \fb_n}}\right|_{t = i\Delta t},
\end{align*}
with $m, n =  1, 2, \ldots, N_e$. The contributions from $\KK$ to the matrices $\QQs_i^{ll}$ and $\QQs_i^{ss}$ respectively read
\begin{align*}
    \MMs_i^{ll} & := \T^{\Sigma H} \GGs^{-1} \KKs_i \PPo^{\Lambda H}, \\
    \MMs_i^{ss} & := \T^\Lambda \GGs^{-1} \PPo^{\Lambda H} \KKs_i \PPo^\Sm.
\end{align*}
Finally, taking into account the contributions from $\TT^\prime$ and $\KK^\prime$, the matrices $\QQs_i^{ll}$ and $\QQs_i^{ss}$ are constructed as follows:
\begin{align*}
    \QQs_i^{ll} & = \T^{\Sigma H} \GGs^{-1}
    \begin{pmatrix}
        \eta \TTs_i + \eta^\prime {\TTs_i}^\prime & - \KKs_i - \KKs_i^\prime \\
        \KKs_i + \KKs_i^\prime & \frac{1}{\eta} \TTs_i + \frac{1}{\eta^\prime} {\TTs_i}^\prime
    \end{pmatrix}
    \PPo^{\Lambda H}, \\
    \QQs_i^{ss} & = \T^\Lambda \GGs^{-1} \PPo^{\Lambda H}
    \begin{pmatrix}
        \eta \TTs_i + \eta^\prime {\TTs_i}^\prime & - \KKs_i - \KKs_i^\prime \\
        \KKs_i + \KKs_i^\prime & \frac{1}{\eta} \TTs_i + \frac{1}{\eta^\prime} {\TTs_i}^\prime
    \end{pmatrix}
    \PPo^\Sm.
\end{align*}

It is noteworthy that $\QQs_i^{ll}$ and $\QQs_i^{ss}$ do not exhibit an infinite tail, since $\TT^s$ and $\KK$ do not contain temporal integration. In particular, $\QQs_i^{ll} = \QQs_i^{ss} = \zrb$ for all $i > k_{\max} := \left\lceil \tfrac{T_{\max}}{\Delta t} \right\rceil$. 

\subsubsection{The matrix $\QQs_i^{ls}$}

Analogous to $\QQs_i^{ll}$ and $\QQs_i^{ss}$, we introduce the following matrices:
\begin{align*}
    \left[\widehat{\TTs}_i\right]_{mn} 
    & := T_{\max} \int_\R \delta_i(t) \inprod{\nv \times \fb_m, \pa_t \TT\paren{q_0(t) \fb_n}} \dt \\
    & \ =  T_{\max} \left.\inprod{\nv \times \fb_m, \pa_t \TT\paren{q_0(t) \fb_n}}\right|_{t = i\Delta t}, \\
    \left[\wh{\KKs}_i\right]_{mn} & := T_{\max} \int_\R \delta_i(t) \inprod{\nv \times \fb_m, \pa_t \KK\paren{q_0(t) \fb_n}} \dt \\
    & \ =  T_{\max} \left.\inprod{\nv \times \fb_m, \pa_t \KK\paren{q_0(t) \fb_n}}\right|_{t = i\Delta t}, 
\end{align*}
with $m, n =  1, 2, \ldots, N_e$. Then, the matrix $\QQs_i^{ls}$ is given by
\[
    \QQs_i^{ls} = \T^{\Sigma H} \GGs^{-1}
    \begin{pmatrix}
        \eta \wh{\TTs}_i + \eta^\prime \wh{\TTs}_i^\prime & - \wh{\KKs}_i - \wh{\KKs}_i^\prime \\
        \wh{\KKs}_i + \wh{\KKs}_i^\prime & \frac{1}{\eta} \wh{\TTs}_i + \frac{1}{\eta^\prime} \wh{\TTs}_i^\prime
    \end{pmatrix}
    \PPo^\Sm.
\]
These matrices also do not exhibit an infinite tail because the temporal integration in $\TT^h$ is canceled out by the temporal differentiation in the rescaling factor $\tilde{\pa}_t$. However, $\QQs_i^{ls} = \zrb$ for all $i > k_{\max} + 1$ rather than $i > k_{\max}$ for $\QQs_i^{ll}$ and $\QQs_i^{ss}$, as the support of $q_0(t)$ exceeds that of $h_0(t)$ by $\Delta t$.

\subsubsection{The matrix $\QQs_i^{sl}$}

Let us denote
\begin{align*}
    \left[\wt{\TTs}_i\right]_{mn} 
    & \ = \dfrac{1}{T_{\max}} \left.\inprod{\nv \times \fb_m, \pa_t^{-1} \TT^s \paren{p_0(t) \fb_n}}\right|_{t = i\Delta t}, \\
    \left[\wt{\KKs}_i\right]_{mn} 
    & \ = \dfrac{1}{T_{\max}} \left.\inprod{\nv \times \fb_m, \pa_t^{-1} \KK\paren{p_0(t) \fb_n}}\right|_{t = i\Delta t}, 
\end{align*}
with $m, n =  1, 2, \ldots, N_e$. The matrix $\QQs_i^{sl}$ is given by
\[
    \QQs_i^{sl} = \T^\Lambda \GGs^{-1} \PPo^{\Lambda H}
    \begin{pmatrix}
        \eta \wt{\TTs}_i + \eta^\prime {\wt{\TTs}_i}^\prime & - \wt{\KKs}_i - \wt{\KKs}_i^\prime \\
        \wt{\KKs}_i + \wt{\KKs}_i^\prime & \frac{1}{\eta} \wt{\TTs}_i + \frac{1}{\eta^\prime} {\wt{\TTs}_i}^\prime
    \end{pmatrix}
    \PPo^{\Lambda H}.
\]
Whereas the matrices $\wt{\TTs}_i$ and $\wt{\TTs}_i^\prime$ do not exhibit an infinite tail (i.e., $\wt{\TTs}_i = \wt{\TTs}_i^\prime = \zrb$ for all $i > k_{\max} - 1$), $\wt{\KKs}_i$ and $\wt{\KKs}_i^\prime$ do because $\pa_t^{-1} \KK$ and $\pa_t^{-1} \KK^\prime$ contain temporal integration. It is now shown that by virtue of the serendipitous interplay between the temporal rescaling operators on one hand, and the Helmholtz splitting on the other, this tail eventually does not lead to an infinite number of non-zero matrices $\QQs_i^{sl}$. Indeed, we note that
\[
    \pa_t^{-1} p_0(t) =  
    \begin{cases}
        0 & t \le -\Delta t, \\
        t + \Delta t \qq & -\Delta t < t \le 0, \\
        \Delta t & \text{otherwise}.
    \end{cases}
\]
It implies the constant tail $\wt{\KKs}_i = \wt{\KKs}_i^\prime = \frac{\Delta t}{T_{\max}} \K_0$ for all $i > k_{\max} - 1$, where
\[
    \left[\K_0\right]_{mn} = \inprod{\nv \times \fb_m, K_0 \fb_n},
\]
with $m, n = 1, 2, \ldots, N_e,$ and the static double-layer operator 
\[
    (K_0\jb)(\xb) = \nv \times p.v. \int_{\Gm} \curlt_{\xxs}  \dfrac{\jb(\yb)}{4\pi \abs{\xb - \yb}} \ds_{\yb}.
\]
In fact, $K_0$ is the limit of the frequency-domain double-layer operator $K_\kappa$ as the wavenumber $\kappa \to 0$. Hence, one can deduce from the large-timestep scaling of the off-diagonal blocks of $\wt{\Qfrak}$ (by setting $\omega = 0$) that
\[
    \T^\Lambda \GGs^{-1} \PPo^{\Lambda H} \K_0 \PPo^{\Lambda H} = \zrb.
\]
This property can also be directly inferred from \cite{CCS+2001}. By combining all properties together, we have 
\begin{equation}
    \label{eq:spec_prop}
    \T^\Lambda \GGs^{-1} \PPo^{\Lambda H} \wt{\KKs}_i \PPo^{\Lambda H} = \T^\Lambda \GGs^{-1} \PPo^{\Lambda H} \wt{\KKs}_i^\prime \PPo^{\Lambda H} = \zrb,
\end{equation}
for all $i > k_{\max} - 1$. Therefore, it follows that $\QQs_i^{sl} = \zrb$ for all $i > k_{\max} - 1$. 

In practice, however, the involved integrals are typically evaluated using quadrature strategies. This approximation introduces discretization errors into the interaction matrices. As a result, the matrices in \eqref{eq:spec_prop} may not vanish. To enforce the theoretical property and eliminate spurious numerical contributions, these matrices should therefore be explicitly set to zero for $i > k_{\max} - 1$. Doing that also reduces computational cost and memory usage. Furthermore, numerical experiments indicate that this truncation improves the stability of the time-domain solution.

\subsection{Computation of Right-Hand Side}
\label{sec:rhs}

The right-hand side vectors in \eqref{eq:fulldis_PMCHWT} are defined by
\begin{equation}
\label{eq:rhs}
    \rrs_i = \int_\R \paren{\dfrac{p_i(t)}{\Delta t} \Pbb^{\Sigma H} + \delta_i(t) \Pbb^\Lambda} \Lbb \T \GGs^{-1}  
    \begin{pmatrix}
        \efrakbm(t) \\
        \hfrakbm(t)
    \end{pmatrix} \dt,
\end{equation}
with $i = 1, 2, \ldots, N_t,$ which consist of the loop contribution
\[
    \rrs_i^l := \dfrac{1}{\Delta t} \T^{\Sigma H} \GGs^{-1} \int_\R p_i(t)  
    \begin{pmatrix}
        \efrakbm(t) \\
        \hfrakbm(t)
    \end{pmatrix} \dt, 
\]
and the star contribution
\begin{align*}
    \rrs_i^s & := \dfrac{1}{T_{\max}} \T^\Lambda \GGs^{-1} \PPo^{\Lambda H} \int_\R \delta_i(t)  
    \begin{pmatrix}
        \pa_t^{-1} \efrakbm(t) \\
        \pa_t^{-1} \hfrakbm(t)
    \end{pmatrix} \dt \\
    & \ = \dfrac{1}{T_{\max}} \T^\Lambda \GGs^{-1} \PPo^{\Lambda H} \left.
    \begin{pmatrix}
        \pa_t^{-1} \efrakbm(t) \\
        \pa_t^{-1} \hfrakbm(t)
    \end{pmatrix} 
    \right|_{t = i\Delta t}.
\end{align*}

Let us assume that the incident transient fields $\eb^{in}$ and $\hb^{in}$ are Gaussian-in-time plane waves of the form
\begin{align}
    \eb^{in}(\xb, t) & = \dfrac{4A}{w \sqrt{\pi}} \pb \exp \paren{-\paren{\dfrac{4}{w}\paren{c(t - t_0) - \kb \cdot \xb}}^2}, \nonumber \\
    \hb^{in}(\xb, t) & = \eta^{-1} \kb \times \eb^{in}(\xb, t), \label{eq:hin}
\end{align}
where $A$ is the amplitude, $\pb$ is the polarization, $\kb$ is the direction, $w$ is the width, and $t_0$ is the time of arrival. Other transient excitations can be treated in an analogous manner.

Because the excitation is Gaussian in time, the incident fields decay exponentially at late times. Consequently, the loop contribution $\rrs_i^l$, which directly follows the transient behavior of the fields, also decays exponentially. In contrast, the star component $\rrs_i^s$ approaches a constant value. This behavior can be understood by noting that the time integral of the Gaussian pulse satisfies
\[
    \dfrac{4A}{w \sqrt{\pi}} \pb \int_\R \exp \paren{-\paren{\dfrac{4}{w}\paren{c(t - t_0) - \kb \cdot \xb}}^2} \dt = \dfrac{A}{c} \pb, 
\]
which yields a finite steady-state contribution. Nevertheless, this limiting constant does not generate a physical field, since the gradient of a spatially constant potential vanishes. Hence, the true steady-state contribution should be zero. In practice, however, numerical errors may prevent $\rrs_i^s$ from decaying completely, leading to spurious late-time residues and potential instability. To ensure stable solutions, the star component should therefore be explicitly forced to zero at late times \cite{ADC+2021}. In time-domain implementations, this can be achieved by truncating $\rrs_i^s$ after the transient has passed, a procedure analogous to the truncation applied to the off-diagonal blocks of $\QQs_i^{sl}$.

In conclusion, \eqref{eq:fulldis_PMCHWT} is the preconditioned and regularized TD-PMCHWT formulation (shorten as the qHP TD-PMCHWT), with the right-hand side vectors defined by \eqref{eq:rhs} and the matrix blocks computed as follows:
\[
    \QQs_i = \QQs_i^{ll} + \QQs_i^{ls} + \QQs_i^{sl} + \QQs_i^{ss}.
\]
Moreover, owing to the finite support of the temporal basis functions and the truncation properties discussed above, the interaction matrices vanish beyond the maximum time lag, i.e., $\QQs_i = \zrb$ for all $i > k_{\max} + 1$.

\subsection{Asymptotic Complexity Analysis}

The proposed formulation involves Calder\'{o}n dense-mesh preconditioning and quasi-Helmholtz regularization, in addition to the assembly of time-domain integral operators. These procedures require the evaluation and multiplication of additional matrices, including $\T_0^s$ and $\T_0^h$, the inversion of the Gram matrix $\GGs$, and the quasi-Helmholtz projectors.

Although the use of BC basis functions increases the assembly time of $\T_0^s, \T_0^h$, and $\GGs$ by constant factors of 36 and 6, respectively, compared with assembly on the original mesh, the asymptotic complexity remains unchanged. In particular, matrix-vector multiplication with $\T_0^s$ and $\T_0^h$ exhibits linear complexity $\OO(N_e)$ when acceleration techniques for static kernels are employed \cite{Bebendorf2000}. The Gram matrix $\GGs$ is sparse and well-conditioned. Therefore, its inversion can be carried out efficiently using either a direct solver (e.g., LU decomposition) or an iterative solver (e.g., GMRES).  Likewise, both the construction of the quasi-Helmholtz projectors and their application require only linear complexity $\OO(N_e)$, since the Moore--Penrose pseudo-inverse appearing in the projectors' definition resembles a graph Laplacian which can be preconditioned with multigrid methods and inverted in linear time \cite{ACB+2013}.

Finally, the quasi-Helmholtz rescaling scheme increases the number of time-domain boundary integral operators to be assembled by a constant factor 3, as described in Section~\ref{sec:computation}. Nevertheless, the formulation can be readily combined with fast time-domain acceleration techniques, such as the plane-wave time-domain (PWTD) method \cite{ESM1999}, the time-domain adaptive integral method \cite{YJM2004}, the nonuniform grid time-domain algorithm \cite{BLM2006}, and accelerated Cartesian expansions \cite{VS2007}. These techniques reduce the computational cost to at worst quasi-linear scaling with respect to both the number of spatial unknowns $N_e$ and temporal unknowns $N_t$. For example, the overall complexity of the proposed scheme is $\OO(N_t N_e \log^2 N_e)$ when coupled with the PWTD method.

\section{Numerical Results}
\label{sec:results}

\begin{figure*}[t]
    \centering
    \includegraphics[trim={35cm 14cm 25cm 16cm}, clip, width=0.22\linewidth]{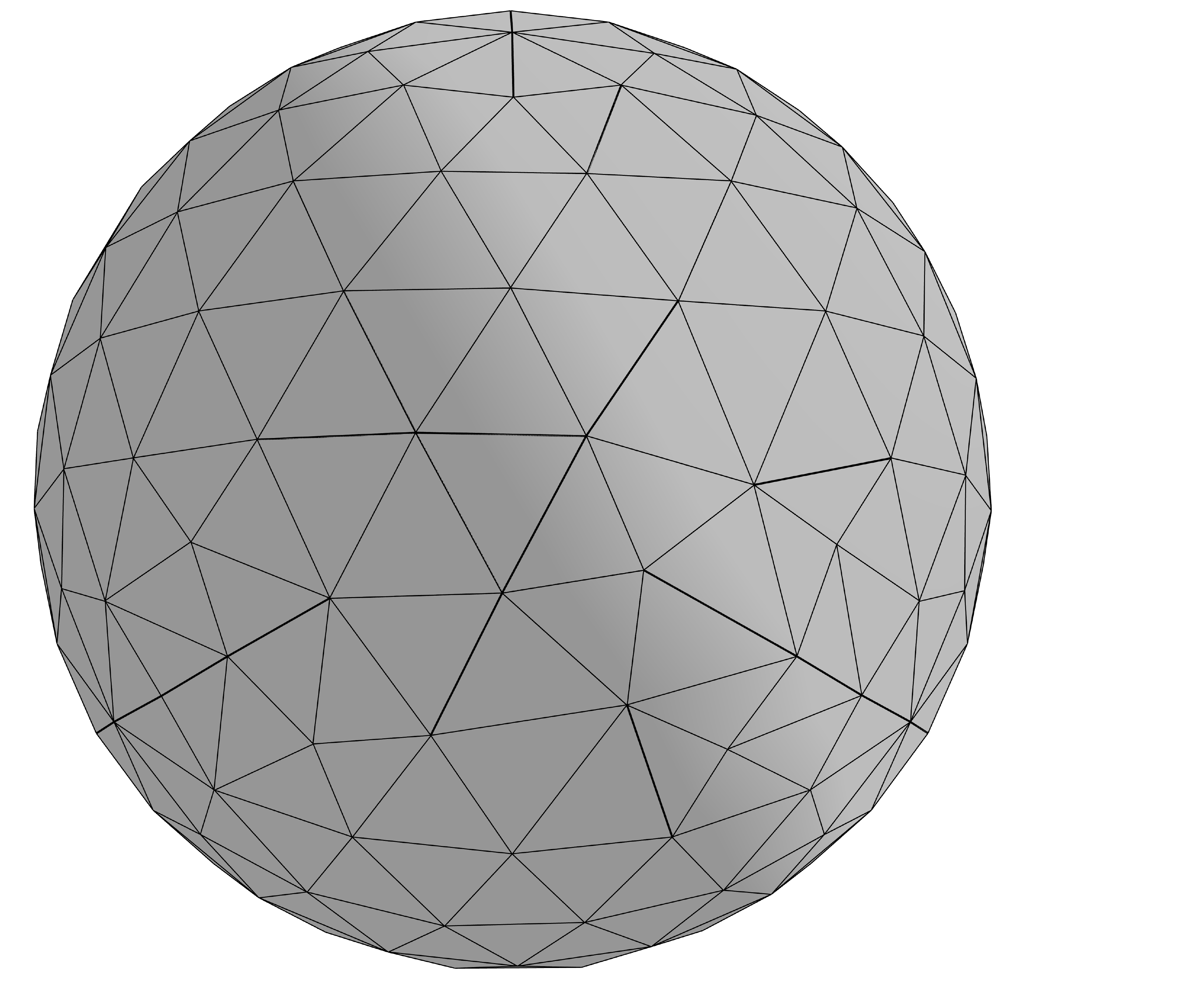}\label{fig:sphere}
    \hfil
    \includegraphics[trim={20cm 10cm 10cm 25cm}, clip, width=0.35\linewidth]{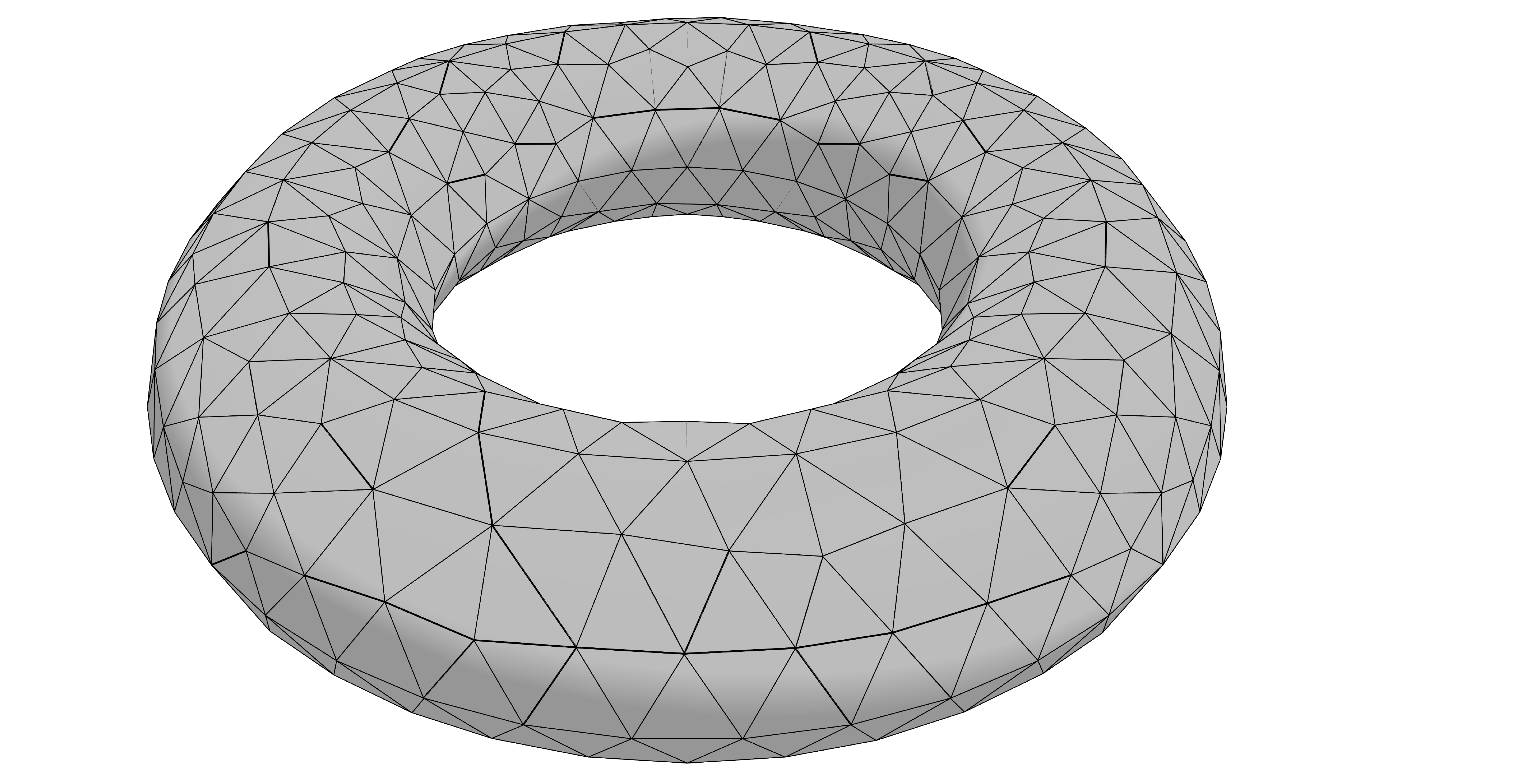}\label{fig:torus}
    \hfil
    \includegraphics[trim={34cm 29cm 34cm 17cm}, clip, width=0.28\textwidth]{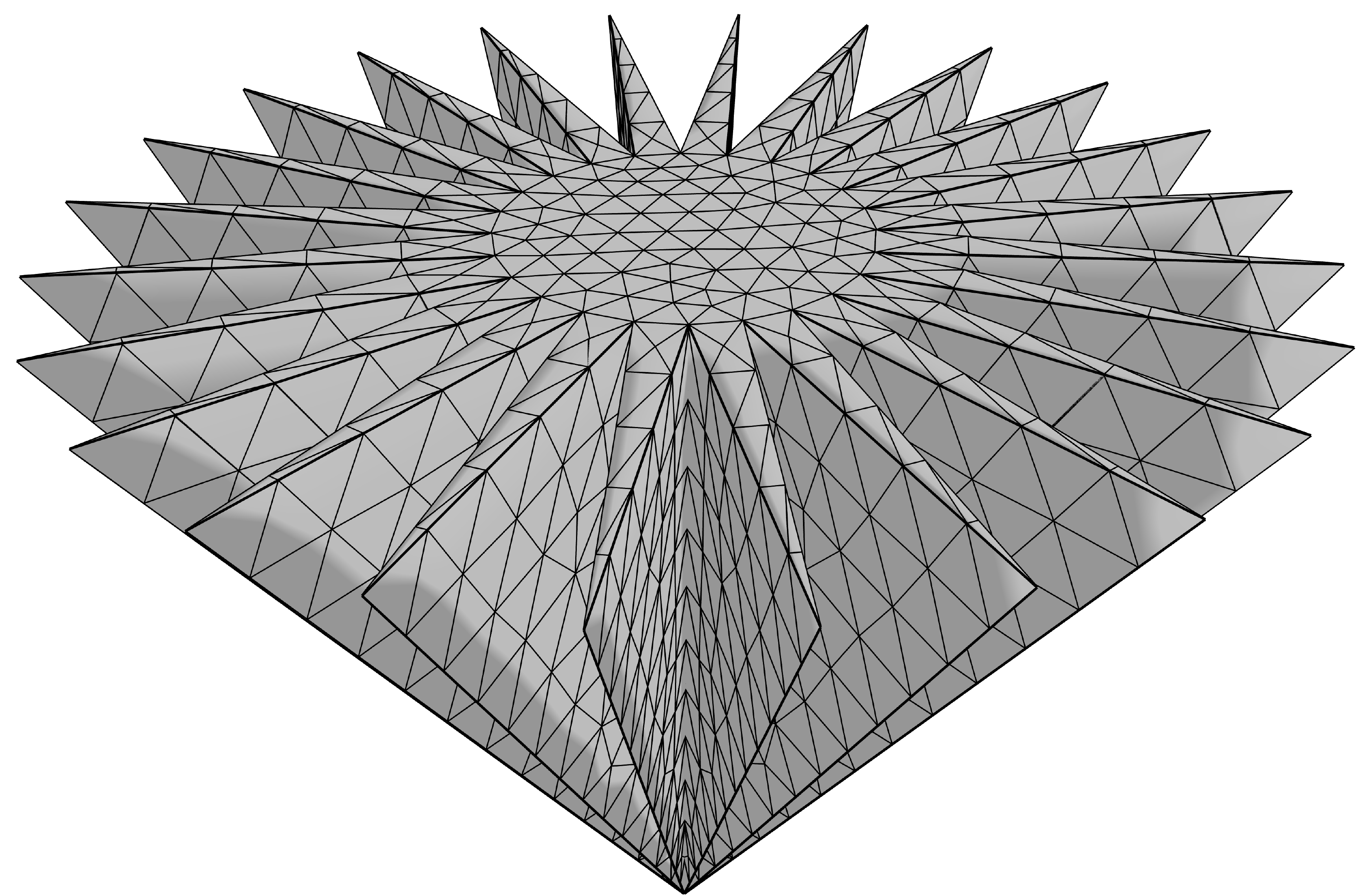}
    \hspace{-0.45cm} 
    \includegraphics[trim={38cm 18cm 38cm 18cm}, clip, width=0.11\textwidth]{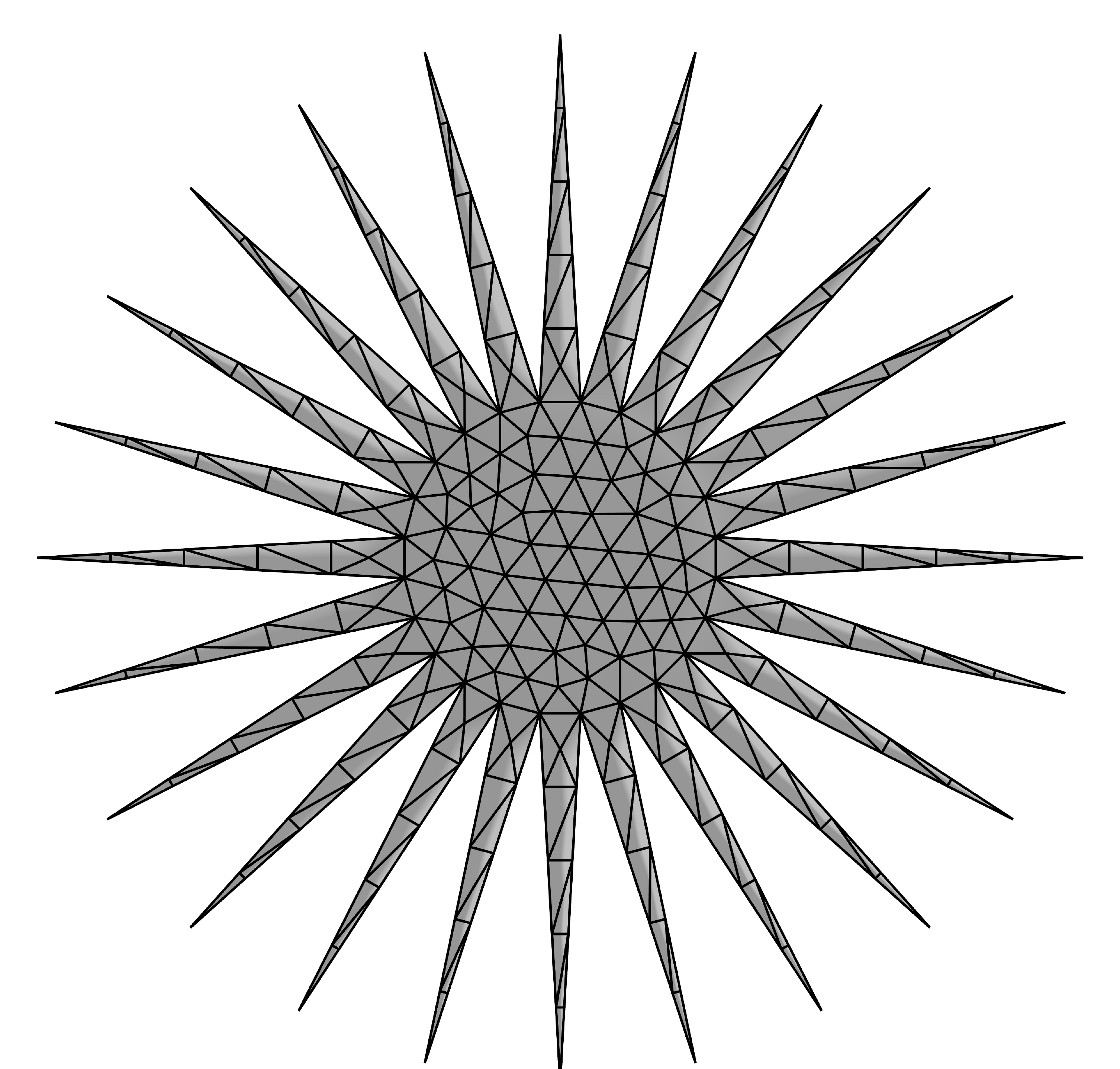} \label{fig:starpyramid}
    \caption{Scatterer geometries used in the numerical experiments. From left to right: a smooth and simply-connected sphere of radius $1\mathrm{m}$; a smooth and multiply-connected torus of 2 radii $0.75\mathrm{m}$ and $0.25\mathrm{m}$; and a highly non-smooth star-based pyramid of height $0.5\mathrm{m}$, with 24 vertices of the base lie on two concentric circles of radius $1\mathrm{m}$ and $0.3\mathrm{m}$.}
\label{fig:geo}
\end{figure*}

In this section, we present some experiments to corroborate the effectiveness of the proposed qHP TD-PMCHWT formulation \eqref{eq:fulldis_PMCHWT}. The numerical examples consider homogeneous dielectric objects of three distinct geometries, which exhibit different topological and smoothness challenges (see Fig.~\ref{fig:geo}):
\begin{itemize}
    \item[(a)] A sphere of radius $1\mathrm{m}$ (smooth and simply-connected);
    \item[(b)] A torus of two radii $0.75\mathrm{m}$ and $0.25\mathrm{m}$ (smooth and multiply-connected);
    \item[(c)] A star-based pyramid of height $0.5\mathrm{m}$ with 24-pointed star base, whose vertices lie on two concentric circles of radius $1\mathrm{m}$ and $0.3\mathrm{m}$ (highly non-smooth) \cite{LC2024b}.
\end{itemize}
In all cases, the background medium is free space, and the excitation is a Gaussian-in-time plane wave defined in \eqref{eq:hin}, with parameters $A = 1\mathrm{V}$, $\pb = (1, 0, 0)^\transpose$, $\kb = (0, 0, 1)^\transpose$, $w = 120 c \Delta t$, and $t_0 = 240 \Delta t$, where the timestep $\Delta t$ will be specified individually for each example. No fast acceleration technique is employed. Instead, a semi-analytic quadrature strategy is used to discretize time-domain boundary integral operators. Specifically, the inner integral is evaluated analytically using the Wilton technique \cite{WRG+1984} and its generalization to retarded-potential interactions proposed by Shanker et al. \cite{SLY+2009}, whereas the outer (testing) integral is computed numerically using $N_q$ quadrature points.

\subsection{Late-Time Stability}

We begin by examining the late-time stability of the proposed qHP TD-PMCHWT formulation through extended versions of the test cases performed in \cite{LGC+2025}.

As the first example, we consider the scattering by a unit dielectric sphere. The unit sphere is approximated by a triangular mesh consisting of 1536 triangles, 770 vertices, and 2304 edges. The interior domain is assumed to be free space, i.e., $(\epsilon^\prime, \mu^\prime) = (\epsilon, \mu)$. In this case, the tangential traces of the incident fields $\paren{\jb, \mb} = \paren{\nv \times \hb^{in}, \eb^{in} \times \nv}$ are the exact solution to the scattering problem. To facilitate comparison with the numerical solution, the exact solution is projected into the RWG space as follows:
\[
    \begin{pmatrix}
        \jjs^{ext}_i \\
        \ms^{ext}_i
    \end{pmatrix}
    =
    - \GGs^{-\transpose}
    \begin{pmatrix}
        \hhs_i \\
        \ees_i
    \end{pmatrix},
\]  
with
\begin{align*}
    \begin{pmatrix}
        \left[\hhs_i\right]_m \\
        \left[\ees_i\right]_m
    \end{pmatrix}
    & = \int_\Gamma \gb_m(\xb) \cdot
    \begin{pmatrix}
        \hb^{in}(\xb, i\Delta t) \\
        - \eb^{in}(\xb, i\Delta t)
    \end{pmatrix}
    \ds,
\end{align*}
for all $i = 1, 2, \ldots, N_t,$ and $m = 1, 2, \ldots, N_e$.
The timestep used in this simulation is $\Delta t =  1.334 \mathrm{ns}$ (or $c \Delta t = 0.4\mathrm{m}$). 

\begin{figure*}[!t]
    \centering
    \includegraphics[trim={0cm 0cm 0.1cm 0.1cm}, clip, width=\linewidth]{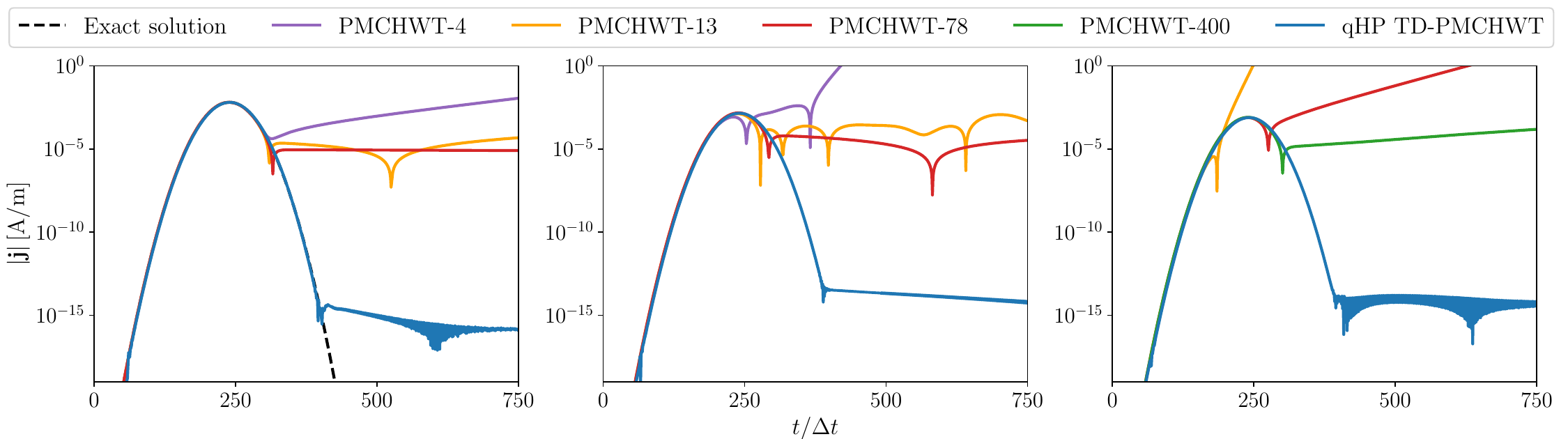}
    \caption{Intensity of the electric current density $\jb$ evaluated at selected points on three scatterers. From left to right: the unit sphere at $\xb = (-0.918, 0.130, -0.363) \mathrm{m}$; the torus at $\xb = (-0.623, -0.416, -0.244) \mathrm{m}$; and the star-based pyramid at $\xb = (0.147, 0.456, 0.050)\mathrm{m}$. Different quadrature strategies are employed to discretize the classical TD-PMCHWT formulation, characterized by the number of quadrature points $N_q$ = 4, 13, 78, and 400 indicated in the legends. Numerical solutions to the TD-PMCHWT equation are severely unstable at late times and highly sensitive to quadrature errors. In contrast, the solution to the proposed qHP TD-PMCHWT formulation is stable even with low-precision integration, using only $N_q = 4$ quadrature points for the sphere and torus and $N_q = 13$ for the star-based pyramid.}
    \label{fig:current}
\end{figure*}

Fig.~\ref{fig:current} \textit{(left)} compares the projected exact solution with numerical solutions to the classical TD-PMCHWT equation \eqref{eq:PMCHWT} and the qHP TD-PMCHWT \eqref{eq:fulldis_PMCHWT}. The classical TD-PMCHWT equation is discretized using a collocation-in-time Galerkin-in-space scheme, as described in \cite{LGC+2025}, with different quadrature strategies employed. As shown in Fig.~\ref{fig:current} \textit{(left)}, the classical formulation suffers from late-time instability, which manifests as non-decaying or exponentially growing errors. This instability is highly sensitive to numerical errors in the computation of the interaction matrices \cite{LGC+2025}. The proposed qHP TD-PMCHWT formulation, in contrast, produces a stable solution that decays to machine precision ($\approx 10^{-15}$ in double precision) at late times. In addition, it closely matches the exact solution. Notably, this stable solution is obtained when using a low-precision quadrature in the integral evaluations.

Next, we investigate scattering by a dielectric torus. The interior domain is filled by a medium with $(\epsilon^\prime, \mu^\prime) = (3\epsilon, \mu)$. The surface $\Gm$ is discretized by 1906 triangles with 953 vertices and 2859 edges, while the time interval is uniformly discretized with a timestep $c\Delta t = 1\mathrm{m}$. Fig.~\ref{fig:current} \textit{(middle)} presents the numerical electric current density $\jb$ computed using the classical TD-PMCHWT and the qHP TD-PMCHWT. Whereas the classical TD-PMCHWT solution becomes unstable at late times, the solution to the qHP TD-PMCHWT is stable. This result confirms the robustness of the proposed formulation for multiply-connected geometries.

The last example involves a highly non-smooth domain, named a star-based pyramid \cite{LC2024b}. This geometry features very sharp corners and edges that significantly amplify numerical quadrature errors in the time-domain discretization \cite{LGC+2025}. The surface is triangulated into 2450 triangles with 1227 vertices and 3675 edges. The pyramid is filled with a medium characterized by  $(\epsilon^\prime, \mu^\prime) = (5\epsilon, \mu)$, and the timestep $c\Delta t = 2\mathrm{m}$. As illustrated in Fig.~\ref{fig:current} \textit{(right)}, numerical solutions to the TD-PMCHWT on this surface are severely unstable, exhibiting very high growth rates even when the integrals are evaluated using extremely accurate quadrature rules. In contrast, the qHP TD-PMCHWT yields a stable solution using only a small number of quadrature points, demonstrating the efficiency of the proposed method.

\begin{figure*}[!t]
    \centering
    \includegraphics[trim={0.1cm 0.1cm 0.7cm 0.3cm}, clip, width=\linewidth]{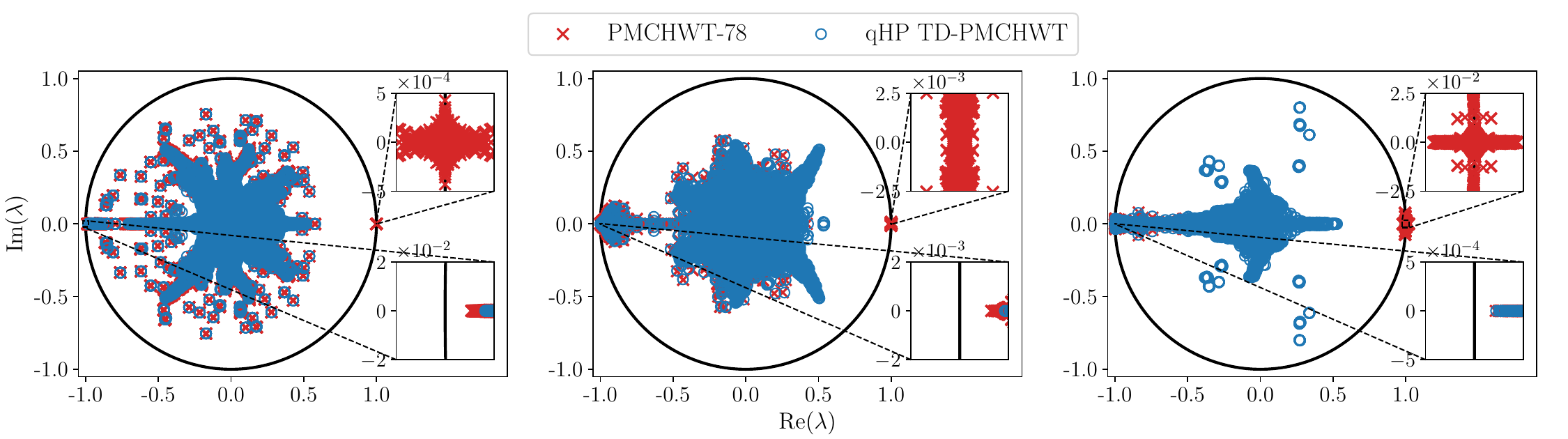}
    \caption{Spectra of the companion matrices for the classical TD-PMCHWT (discretized using $N_q = 78$ quadrature points) and the proposed qHP TD-PMCHWT on the unit sphere, the toroidal surface, and the star-based pyramid (from left to right). For the qHP TD-PMCHWT, $N_q = 4$ is used for the sphere and torus, and $N_q = 13$ for the star-based pyramid. The TD-PMCHWT's spectra exhibit eigenvalues clustered around $1+0\iota$, some of which lie outside the unit circle and lead to late-time instability in the numerical solution. In contrast, the qHP TD-PMCHWT does not support any eigenvalue around $1+0\iota$, rendering its solution stable. In all cases, the eigenvalues near $-1+0\iota$ remain strictly inside the unit circle and correspond to oscillations below machine precision \cite{VVV+2013}.}
    \label{fig:eigen}
\end{figure*}

The stability behavior of the classical TD-PMCHWT and the proposed qHP TD-PMCHWT formulations is further examined by means of companion matrix stability analysis. This technique was first proposed in \cite{DWB1998}, and recently extended to time-domain integral equation systems with constant infinite tails (e.g., TD-EFIE and TD-PMCHWT) in \cite{VDZ+2022,LGC+2025}. Fig.~\ref{fig:eigen} shows the spectra of the companion matrices for both formulations in the three examples considered above. The companion matrices of the classical TD-PMCHWT exhibit eigenvalues clustered around $1+0\iota$, with several shifted outside the unit circle due to numerical errors, causing instability in the time-domain solution. In contrast, all eigenvalues of the qHP TD-PMCHWT reside strictly inside the unit circle, with none located near $1 + 0\iota$, indicating that the qHP TD-PMCHWT is immune to late-time instability.

It is worth noting that the late-time stability of the proposed formulation holds for all timestep sizes. In the three examples above, moderate timesteps were chosen for demonstration purposes. For simulations in the high-frequency regime (corresponding to small timesteps), denser meshes and more accurate quadrature rules are typically required to properly resolve the oscillatory behavior of the electromagnetic fields. Under these conditions, fast acceleration techniques become essential to mitigate the increased computational cost. Consequently, the robustness of the qHP TD-PMCHWT formulation against numerical quadrature errors becomes even more important, since such acceleration and approximation methods inevitably introduce errors of a nature similar to quadrature errors.

\subsection{Conditioning}

\begin{figure*}[!t]
    \centering
    \includegraphics[trim={0.2cm 0cm 0.2cm 0.1cm}, clip, width=\linewidth]{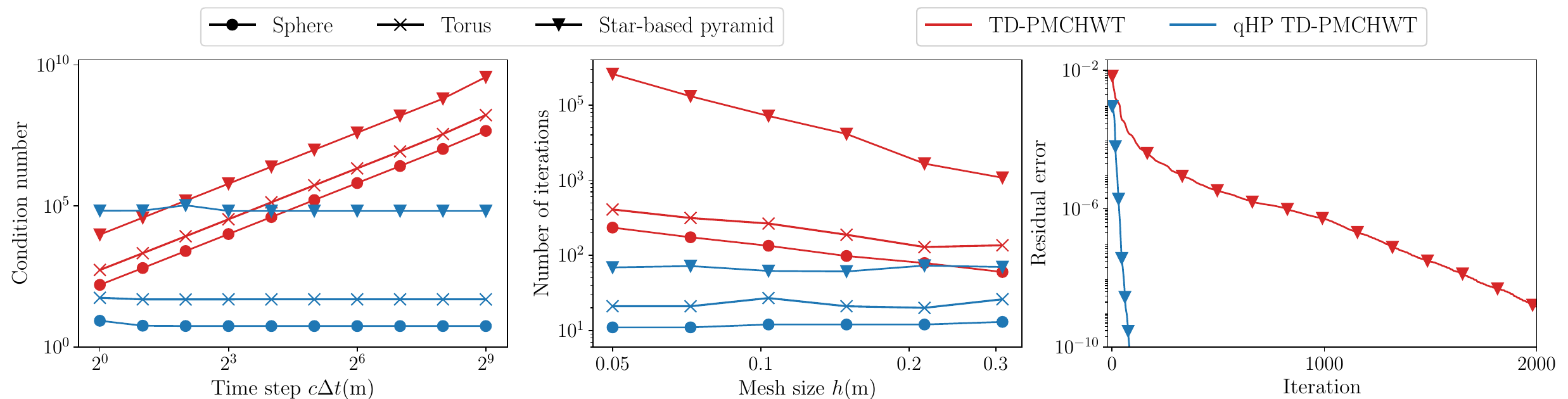}
    \caption{\textit{(Left)} Condition number of the classical TD-PMCHWT and qHP TD-PMCHWT systems as a function of the timestep $\Delta t$. \textit{(Middle)} Number of GMRES iterations required to reach a relative tolerance of $10^{-6}$ for the TD-PMCHWT and qHP TD-PMCHWT formulations as a function of the mesh size $h$. \textit{(Right)} GMRES convergence history for the star-based pyramid scattering problem with mesh size $h = 0.08\mathrm{m}$ and timestep $c\Delta t = 5\mathrm{m}$. The classical TD-PMCHWT formulation suffers from both dense-mesh breakdown and large-timestep breakdown, whereas the proposed qHP TD-PMCHWT formulation is immune to both phenomena.}
    \label{fig:cond}
\end{figure*}

We continue by examining the conditioning properties of the proposed formulation. First, the average mesh size is fixed at $h = 0.3\mathrm{m}$ for the sphere, $h = 0.15\mathrm{m}$ for the torus, and $h = 0.16\mathrm{m}$ for the star-based pyramid. The condition number of the matrix $\QQs_0$, which appears on the left-hand side of the MOT system \eqref{eq:MOT} that needs to be solved at each time step, is computed with timesteps $c \Delta t$ ranging from $1 \mathrm{m}$ to $512\mathrm{m}$. As illustrated in Fig.~\ref{fig:cond} \textit{(left)}, the condition number of the classical TD-PMCHWT system grows quadratically with $\Delta t$, whereas that of the qHP TD-PMCHWT remains essentially constant. 

Next, we fix the timestep to $c\Delta t = 1\mathrm{m}$ for the sphere, $2 \mathrm{m}$ for the torus, and $5\mathrm{m}$ for the star-based pyramid, while decreasing $h$ from $0.31\mathrm{m}$ to $0.05\mathrm{m}$. The number of GMRES iterations required to reach a relative tolerance of $10^{-6}$ when solving system \eqref{eq:MOT} for the right-hand side $\rrs_i$ at time step $i = 250$ is presented in Fig.~\ref{fig:cond} \textit{(middle)} as a function of $h$. This time step is chosen to lie near the peak of the Gaussian pulse. The results show that the qHP TD-PMCHWT maintains a stable iteration count under mesh refinement, in contrast to the classical formulation whose iteration count grows with mesh density.

To corroborate the practical effectiveness of the preconditioning schemes, we further report the GMRES convergence histories for the scattering problem on the star-based pyramid in Fig.~\ref{fig:cond} \textit{(right)}. In this experiment, the mesh size is $h = 0.08\mathrm{m}$ and the timestep is $c\Delta t = 5\mathrm{m}$. The TD-PMCHWT system fails to reach a residual error of $10^{-6}$ within 800 iterations, whereas the qHP TD-PMCHWT converges to a residual of $10^{-8}$ in only 56 iterations. These results demonstrate that the proposed qHP TD-PMCHWT formulation yields rapidly convergent linear systems, thereby enabling efficient solution of the MOT scheme with iterative solvers.

\subsection{Far-Field Computation}

In the final experiment, we assess the accuracy of the scattered far fields obtained from the qHP TD-PMCHWT solution in both full-wave and low-frequency regimes. The unit sphere, torus, and star-based pyramid are discretized with average mesh sizes $h = 0.2\mathrm{m}, 0.13\mathrm{m}$, and $0.16\mathrm{m}$, respectively. The surface current densities $\jb$ and $\mb$ are computed with timesteps $\Delta t = 0.05 \mathrm{ns}$ (equivalently, $c\Delta t = 1.499 \times 10^{-2} \mathrm{m}$) and $\Delta t = 100 \mathrm{s}$ ($c\Delta t = 2.998 \times 10^{10} \mathrm{m}$), and subsequently Fourier transformed at $200\mathrm{MHz}$ and $10^{-5} \mathrm{Hz}$, respectively.

\begin{figure*}[!t]
    \centering
    \includegraphics[trim={0cm 0cm 0.1cm 0.1cm}, clip, width=\linewidth]{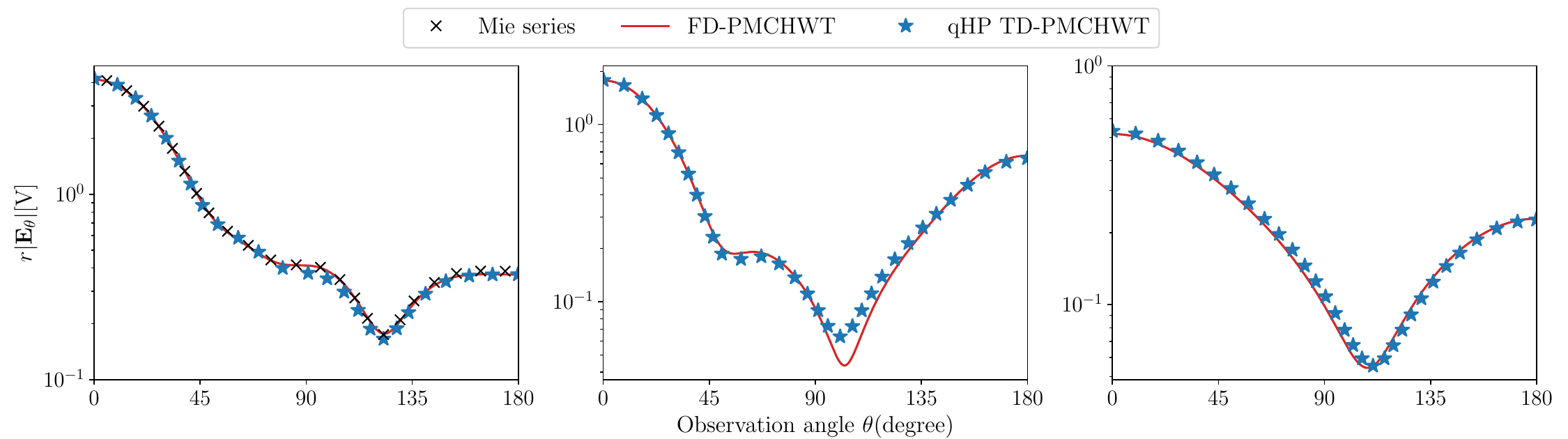}
    \caption{Scattered electric far field of an incident plane wave at the frequency $200 \mathrm{MHz}$, computed using the Mie series, FD-PMCHWT, and qHP TD-PMCHWT  formulations for the unit sphere, torus, and star-based pyramid. The far fields obtained from the qHP TD-PMCHWT solution closely match those obtained from the FD-PMCHWT and the reference Mie-series solution.}
    \label{fig:farfield_highfreq}
\end{figure*}

Fig.~\ref{fig:farfield_highfreq} shows the scattered electric far fields derived from the numerical surface currents at $200 \mathrm{MHz}$ obtained with the qHP TD-PMCHWT formulation, which closely match those computed from the standard FD-PMCHWT formulation. For the sphere, the results are further validated by comparison with the exact Mie-series solution \cite{HRA2023}. 

\begin{figure*}[!t]
    \centering
    \includegraphics[trim={0cm 0cm 0.1cm 0.1cm}, clip, width=\linewidth]{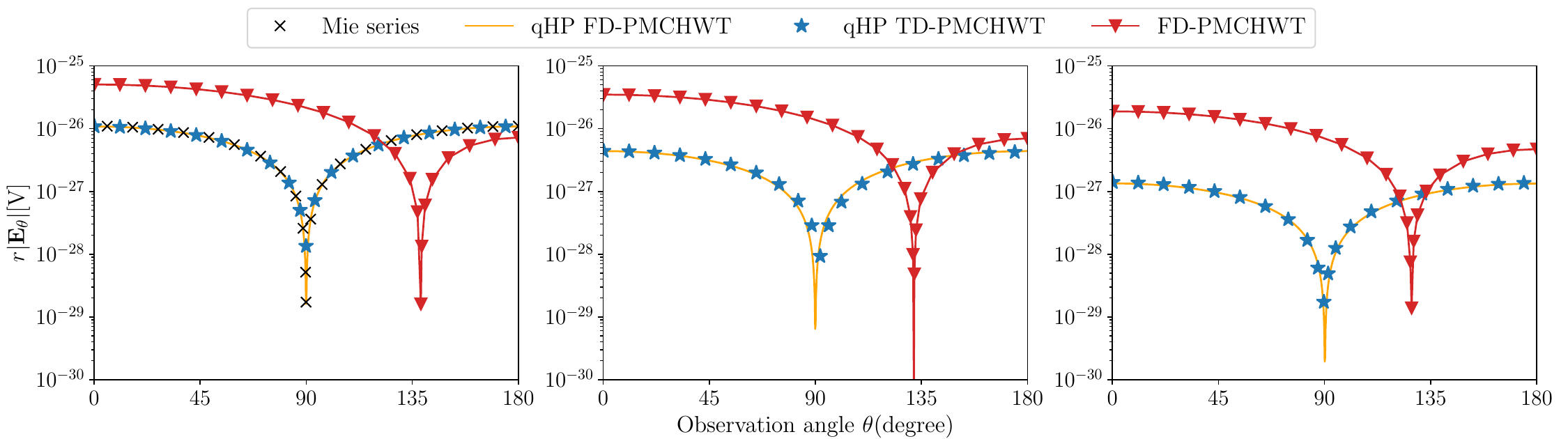}
    \caption{Scattered electric far field of an incident plane wave at the frequency $10^{-5} \mathrm{Hz}$, computed using the Mie series, FD-PMCHWT, qHP FD-PMCHWT, and qHP TD-PMCHWT formulations for the unit sphere, torus, and star-based pyramid. The far fields obtained from the qHP TD-PMCHWT and qHP FD-PMCHWT formulations are accurate and in close agreement with the reference Mie-series solution, whereas the classical FD-PMCHWT result is highly inaccurate due to loss of numerical accuracy at low frequencies.}
    \label{fig:farfield}
\end{figure*}

In the low-frequency regime, the far fields computed using the FD-PMCHWT formulation become highly inaccurate due to numerical cancellation in the surface currents. In this regime, where the wavelength is much larger than the characteristic diameter of the scatterer, the far-field responses of different geometries exhibit similar radiation patterns with symmetry about $90^\circ$. As illustrated in Fig.~\ref{fig:farfield}, the far fields obtained from the qHP TD-PMCHWT formulation accurately capture this behavior and are in close agreement with those computed using its frequency-domain counterpart introduced in \cite{BMC+2017} and the Mie-series solution.

These results confirm the accuracy of the derived quantities obtained with the proposed formulation at moderately small timesteps, as well as its significantly improved accuracy in the large-timestep regime. A direct comparison with the classical TD-PMCHWT formulation is not meaningful in either regime, as its time-domain solutions fail to decay at late times, leading to unreliable frequency-domain surface currents after Fourier transformation.

\section{Conclusion}
\label{sec:conclusion}

We have proposed a new time-domain formulation for electromagnetic scattering by homogeneous dielectric objects, building upon the TD-PMCHWT equation and the loop-star decomposition via quasi-Helmholtz projectors. A multiplicative diagonal preconditioner, based on a modified static EFIO, was introduced to eliminate dense-mesh breakdown. Ill-conditioning and loss of accuracy in the derived quantities at large timesteps were simultaneously mitigated through a rescaling procedure, while incorporating temporal integration and differentiation as rescaling factors effectively suppressed the notorious late-time instability. An appropriate temporal testing scheme was applied, with particular attention to practical implementation. Numerical experiments across various scenarios confirmed the stability and robustness of the proposed formulation, as well as the improved accuracy in the far-field evaluation in the large-timestep regime. The robustness against numerical quadrature errors becomes even more critical when employing compression techniques, such as PWTD algorithms, as they unavoidably introduce errors that are similar to quadrature errors.

Future research may explore advanced discretization schemes to further enhance performance, such as higher-order spatial and temporal basis functions or time discretization based on convolution quadrature. 


%




\ifCLASSOPTIONcaptionsoff
  \newpage
\fi



\bibliographystyle{IEEEtran}
\bibliography{abrv_ref.bib}

\begin{thebibliography}{10}
\providecommand{\url}[1]{#1}
\csname url@samestyle\endcsname
\providecommand{\newblock}{\relax}
\providecommand{\bibinfo}[2]{#2}
\providecommand{\BIBentrySTDinterwordspacing}{\spaceskip=0pt\relax}
\providecommand{\BIBentryALTinterwordstretchfactor}{4}
\providecommand{\BIBentryALTinterwordspacing}{\spaceskip=\fontdimen2\font plus
\BIBentryALTinterwordstretchfactor\fontdimen3\font minus \fontdimen4\font\relax}
\providecommand{\BIBforeignlanguage}[2]{{%
\expandafter\ifx\csname l@#1\endcsname\relax
\typeout{** WARNING: IEEEtran.bst: No hyphenation pattern has been}%
\typeout{** loaded for the language `#1'. Using the pattern for}%
\typeout{** the default language instead.}%
\else
\language=\csname l@#1\endcsname
\fi
#2}}
\providecommand{\BIBdecl}{\relax}
\BIBdecl

\bibitem{PM1973}
A.~J. Poggio and E.~K. Miller, \emph{Integral equation solutions of three-dimensional scattering problems}.\hskip 1em plus 0.5em minus 0.4em\relax Elsevier Science \& Technology, 1973, pp. 159--264.

\bibitem{Muller1969}
C.~M\"{u}ller, \emph{Foundations of the mathematical theory of electromagnetic waves}.\hskip 1em plus 0.5em minus 0.4em\relax Springer Berlin Heidelberg, 1969.

\bibitem{YTJ2008}
P.~Yl\"{a}-Oijala, M.~Taskinen, and S.~J\"{a}rvenp\"{a}\"{a}, ``Analysis of surface integral equations in electromagnetic scattering and radiation problems,'' \emph{Eng. Anal. Bound. Elem.}, vol.~32, no.~3, pp. 196--209, 2008.

\bibitem{YJN2013}
S.~Yan, J.-M. Jin, and Z.~Nie, ``Accuracy improvement of the second-kind integral equations for generally shaped objects,'' \emph{IEEE Trans. Antennas Propag.}, vol.~61, no.~2, pp. 788--797, 2013.

\bibitem{Hiptmair2006}
R.~Hiptmair, ``{Operator preconditioning},'' \emph{Comput. Math. Appl.}, vol.~52, no.~5, pp. 699--706, 2006.

\bibitem{YJN2010a}
S.~Yan, J.-M. Jin, and Z.~Nie, ``{A comparative study of Calder\'{o}n preconditioners for PMCHWT equations},'' \emph{IEEE Trans. Antennas Propag.}, vol.~58, no.~7, pp. 2375--2383, 2010.

\bibitem{CAM2011}
K.~Cools, F.~P. Andriulli, and E.~Michielssen, ``{A Calder\'{o}n multiplicative preconditioner for the PMCHWT integral equation},'' \emph{{IEEE} Trans. Antennas Propag.}, vol.~59, no.~12, pp. 4579--4587, 2011.

\bibitem{NN2012}
K.~Niino and N.~Nishimura, ``{Calder\'{o}n preconditioning approaches for PMCHWT formulations for Maxwell's equations},'' \emph{Int. J. Numer. Model.: Electron. Netw. Devices Fields}, vol.~25, pp. 558--572, 2012.

\bibitem{KBH+2022}
A.~Kleanthous, T.~Betcke, D.~P. Hewett, P.~Escapil-Inchausp\'{e}, C.~Jerez-Hanckes, and A.~J. Baran, ``{Accelerated Calder\'{o}n preconditioning for Maxwell transmission problems},'' \emph{J. Comput. Phys.}, vol. 458, p. 111099, 2022.

\bibitem{LCA+2023}
V.~C. Le, P.~Cordel, F.~P. Andriulli, and K.~Cools, ``{A Yukawa-Calder\'{o}n time-domain combined field integral equation for electromagnetic scattering},'' in \emph{Proc. Int. Conf. Electromagn. Adv. Appl. (ICEAA)}, 2023.

\bibitem{LCA+2024}
------, ``{A stabilized time-domain combined field integral equation using the quasi-Helmholtz projectors},'' \emph{IEEE Trans. Antennas Propag.}, vol.~72, no.~7, pp. 5852--5864, 2024.

\bibitem{CAO+2009b}
K.~Cools, F.~P. Andriulli, F.~Olyslager, and E.~Michielssen, ``{Time domain Calder{\'{o}}n identities and their application to the integral equation analysis of scattering by {PEC} objects Part I: Preconditioning},'' \emph{{IEEE} Trans. Antennas Propag.}, vol.~57, no.~8, pp. 2352--2364, 2009.

\bibitem{ACB+2013}
F.~P. Andriulli, K.~Cools, I.~Bogaert, and E.~Michielssen, ``{On a well-conditioned electric field integral operator for multiply connected geometries},'' \emph{{IEEE} Trans. Antennas Propag.}, vol.~61, no.~4, pp. 2077--2087, 2013.

\bibitem{BMC+2017}
Y.~Beghein, R.~Mitharwal, K.~Cools, and F.~P. Andriulli, ``{On a low-frequency and refinement stable PMCHWT integral equation leveraging the quasi-Helmholtz projectors},'' \emph{{IEEE} Trans. Antennas Propag.}, vol.~65, no.~10, pp. 5365--5375, 2017.

\bibitem{GAM+2017}
J.~E.~O. Guzman, S.~B. Adrian, R.~Mitharwal, Y.~Beghein, T.~F. Eibert, K.~Cools, and F.~P. Andriulli, ``{On the hierarchical preconditioning of the {PMCHWT} integral equation on simply and multiply connected geometries},'' \emph{{IEEE} Antennas Wirel. Propag. Lett.}, vol.~16, pp. 1044--1047, 2017.

\bibitem{MBC+2020}
A.~Merlini, Y.~Beghein, K.~Cools, E.~Michielssen, and F.~P. Andriulli, ``{Magnetic and combined field integral equations based on the quasi-Helmholtz projectors},'' \emph{{IEEE} Trans. Antennas Propag.}, vol.~68, no.~5, pp. 3834--3846, 2020.

\bibitem{GSM+2025}
V.~Giunzioni, A.~Scazzola, A.~Merlini, and F.~P. Andriulli, ``{Low-frequency stabilizations of the PMCHWT equation for dielectric and conductive media: On a full-wave alternative to eddy-current solvers},'' \emph{IEEE Trans. Antennas Propag.}, vol.~73, no.~8, pp. 5725--5740, 2025.

\bibitem{BCA2015}
Y.~Beghein, K.~Cools, and F.~P. Andriulli, ``{A DC stable and large-time step well-balanced TD-EFIE based on quasi-Helmholtz projectors},'' \emph{{IEEE} Trans. Antennas Propag.}, vol.~63, no.~7, pp. 3087--3097, 2015.

\bibitem{BCA2015b}
------, ``{A DC-stable, well-balanced, Calder{\'{o}}n preconditioned time domain electric field integral equation},'' \emph{{IEEE} Trans. Antennas Propag.}, vol.~63, no.~12, pp. 5650--5660, 2015.

\bibitem{BCA2015d}
------, ``A robust and low frequency stable time domain {PMCHWT} equation,'' in \emph{Proc. Int. Conf. Electromagn. Adv. Appl. (ICEAA)}, 2015, pp. 954--957.

\bibitem{DAC2020}
A.~Dely, F.~P. Andriulli, and K.~Cools, ``Large time step and {DC} stable {TD}-{EFIE} discretized with implicit {Runge-Kutta} methods,'' \emph{{IEEE} Trans. Antennas Propag.}, vol.~68, no.~2, pp. 976--985, 2020.

\bibitem{MB1982}
H.~Mieras and C.~Bennett, ``Space-time integral equation approach to dielectric targets,'' \emph{IEEE Trans. Antennas Propag.}, vol.~30, no.~1, pp. 2--9, 1982.

\bibitem{LGC+2025}
V.~C. Le, V.~Giunzioni, P.~Cordel, F.~P. Andriulli, and K.~Cools, ``On the late-time instability of {MOT} solution to the time-domain {PMCHWT} equation,'' \emph{{IEEE} Antennas Wirel. Propag. Lett.}, vol.~24, no.~12, pp. 4720--4724, 2025.

\bibitem{SA1993}
A.~Sadigh and E.~Arvas, ``Treating the instabilities in marching-on-in-time method from a different perspective (electromagnetic scattering),'' \emph{IEEE Trans. Antennas Propag.}, vol.~41, no.~12, pp. 1695--1702, 1993.

\bibitem{WPC+2004}
D.~S. Weile, G.~Pisharody, N.-W. Chen, B.~Shanker, and E.~Michielssen, ``{A novel scheme for the solution of the time-domain integral equations of electromagnetics},'' \emph{{IEEE} Trans. Antennas Propag.}, vol.~52, no.~1, pp. 283--295, 2004.

\bibitem{SLY+2009}
B.~Shanker, M.~Lu, J.~Yuan, and E.~Michielssen, ``{Time domain integral equation analysis of scattering from composite bodies via exact evaluation of radiation fields},'' \emph{{IEEE} Trans. Antennas Propag.}, vol.~57, no.~5, pp. 1506--1520, 2009.

\bibitem{YE2006}
A.~C. Yucel and A.~A. Ergin, ``{Exact evaluation of retarded-time potential integrals for the RWG bases},'' \emph{IEEE Trans. Antennas Propag.}, vol.~54, no.~5, pp. 1496--1502, 2006.

\bibitem{VVV+2013}
E.~van~’t Wout, D.~R. van~der Heul, H.~van~der Ven, and C.~Vuik, ``The influence of the exact evaluation of radiation fields in finite precision arithmetic on the stability of the time domain integral equation method,'' \emph{IEEE Trans. Antennas Propag.}, vol.~61, no.~12, pp. 6064--6074, 2013.

\bibitem{TXX2014}
X.~Tian, G.~Xiao, and S.~Xiang, ``Application of analytical expressions for retarded-time potentials in analyzing the transient scattering by dielectric objects,'' \emph{IEEE Antennas Wirel. Propag. Lett.}, vol.~13, pp. 1313--1316, 2014.

\bibitem{LAC2024}
V.~C. Le, F.~P. Andriulli, and K.~Cools, ``{A DC stable, well-conditioned and low-frequency regularized time-domain PMCHWT equation},'' in \emph{Proc. 2024 {IEEE} Int. Symp. Antennas Propag. and {USNC}-{URSI} Radio Sci. Meet.}, 2024.

\bibitem{ZC2000}
J.-S. Zhao and W.~C. Chew, ``{Integral equation solution of Maxwell's equations from zero frequency to microwave frequencies},'' \emph{{IEEE} Trans. Antennas Propag.}, vol.~48, no.~10, pp. 1635--1645, 2000.

\bibitem{CCS+2001}
S.~Y. Chen, W.~C. Chew, J.~M. Song, and J.-S. Zhao, ``Analysis of low frequency scattering from penetrable scatterers,'' \emph{{IEEE} Trans. Geosci. Remote Sens.}, vol.~39, no.~4, pp. 726--735, 2001.

\bibitem{BCA+2014}
I.~Bogaert, K.~Cools, F.~P. Andriulli, and H.~Bagci, ``{Low-frequency scaling of the standard and mixed magnetic field and Müller integral equations},'' \emph{{IEEE} Trans. Antennas Propag.}, vol.~62, no.~2, pp. 822--831, 2014.

\bibitem{Andriulli2012}
F.~P. Andriulli, ``{Loop-star and loop-tree decompositions: Analysis and efficient algorithms},'' \emph{{IEEE} Trans. Antennas Propag.}, vol.~60, no.~5, pp. 2347--2356, 2012.

\bibitem{Beghein2015}
Y.~Beghein, ``{Advanced discretization and preconditioning techniques for electromagnetic boundary integral equations},'' Ph.D. dissertation, Ghent University. Faculty of Engineering and Architecture, Ghent, Belgium, 2015.

\bibitem{RWG1982}
S.~Rao, D.~Wilton, and A.~Glisson, ``Electromagnetic scattering by surfaces of arbitrary shape,'' \emph{{IEEE} Trans. Antennas Propag.}, vol. AP-30, no.~3, pp. 409--418, 1982.

\bibitem{HS2003b}
R.~Hiptmair and C.~Schwab, ``{Natural boundary element methods for the electric field integral equation on polyhedra},'' \emph{{SIAM} J. Numer. Anal.}, vol.~40, no.~1, pp. 66--86, 2003.

\bibitem{BC2007}
A.~Buffa and S.~H. Christiansen, ``A dual finite element complex on the barycentric refinement,'' \emph{Math. Comp.}, vol.~76, no. 260, pp. 1743--1770, 2007.

\bibitem{Buffa2001}
A.~Buffa, ``Hodge decompositions on the boundary of nonsmooth domains: the multi-connected case,'' \emph{Math. Models Methods Appl. Sci.}, vol.~11, no.~09, pp. 1491--1503, 2001.

\bibitem{BC2001b}
A.~Buffa and P.~Ciarlet, ``{On traces for functional spaces related to Maxwell's equations Part {II}: Hodge decompositions on the boundary of Lipschitz polyhedra and applications},'' \emph{Math. Methods Appl. Sci.}, vol.~24, pp. 31--48, 2001.

\bibitem{BHV+2003}
A.~Buffa, R.~Hiptmair, T.~von Petersdorff, and C.~Schwab, ``{Boundary element methods for Maxwell transmission problems in Lipschitz domains},'' \emph{Numer. Math.}, vol.~95, no.~3, pp. 459--485, 2003.

\bibitem{LC2024}
V.~C. Le and K.~Cools, ``An operator preconditioned combined field integral equation for electromagnetic scattering,'' \emph{SIAM J. Numer. Anal.}, vol.~62, no.~6, pp. 2484--2505, 2024.

\bibitem{HEA+2023}
B.~Hofmann, T.~F. Eibert, F.~P. Andriulli, and S.~B. Adrian, ``An excitation-aware and self-adaptive frequency normalization for low-frequency stabilized electric field integral equation formulations,'' \emph{{IEEE} Trans. Antennas Propag.}, vol.~71, no.~5, pp. 4301--4314, 2023.

\bibitem{ADC+2021}
S.~B. Adrian, A.~Dely, D.~Consoli, A.~Merlini, and F.~P. Andriulli, ``Electromagnetic integral equations: insights in conditioning and preconditioning,'' \emph{IEEE Open J. Antennas Propag.}, vol.~2, pp. 1143--1174, 2021.

\bibitem{Bebendorf2000}
M.~Bebendorf, ``Approximation of boundary element matrices,'' \emph{Numer. Math.}, vol.~86, no.~4, pp. 565--589, 2000.

\bibitem{ESM1999}
A.~A. Ergin, B.~Shanker, and E.~Michielssen, ``The plane-wave time-domain algorithm for the fast analysis of transient wave phenomena,'' \emph{{IEEE} Antennas Propag. Mag.}, vol.~41, no.~4, pp. 39--52, 1999.

\bibitem{YJM2004}
A.~E. Yilmaz, J.-M. Jin, and E.~Michielssen, ``Time domain adaptive integral method for surface integral equations,'' \emph{{IEEE} Trans. Antennas Propag.}, vol.~52, no.~10, pp. 2692--2708, 2004.

\bibitem{BLM2006}
A.~Boag, V.~Lomakin, and E.~Michielssen, ``{Nonuniform grid time domain (NGTD) algorithm for fast evaluation of transient wave fields},'' \emph{{IEEE} Trans. Antennas Propag.}, vol.~54, no.~7, pp. 1943--1951, 2006.

\bibitem{VS2007}
M.~Vikram and B.~Shanker, ``{Fast evaluation of time domain fields in sub-wavelength source/observer distributions using accelerated Cartesian expansions (ACE)},'' \emph{J. Comput. Phys.}, vol. 227, no.~2, pp. 1007--1023, 2007.

\bibitem{LC2024b}
V.~C. Le and K.~Cools, ``Boundary element methods for the magnetic field integral equation on polyhedra,'' \emph{preprint}, 2024.

\bibitem{WRG+1984}
D.~Wilton, S.~Rao, A.~Glisson, D.~Schaubert, O.~Al-Bundak, and C.~Butler, ``Potential integrals for uniform and linear source distributions on polygonal and polyhedral domains,'' \emph{IEEE Trans. Antennas Propag.}, vol.~32, no.~3, pp. 276--281, 1984.

\bibitem{DWB1998}
S.~J. Dodson, S.~P. Walker, and M.~J. Bluck, ``Implicitness and stability of time domain integral equation scattering analyses,'' \emph{Appl. Comput. Electromagn. Soc. J.}, vol.~13, no.~3, pp. 291--301, 1998.

\bibitem{VDZ+2022}
P.~Van~Diepen, R.~J. Dilz, P.~Zwamborn, and M.~C. Van~Beurden, ``{The role of Jordan blocks in the MOT-scheme time domain EFIE linear-in-time solution instability},'' \emph{Prog. Electromagn. Res.}, vol.~95, pp. 123--140, 2022.

\bibitem{HRA2023}
B.~Hofmann, P.~Respondek, and S.~B. Adrian, ``{SphericalScattering: A Julia package for electromagnetic scattering from spherical objects},'' \emph{J. Open Source Softw.}, vol.~8, no.~91, p. 5820, 2023.

\end{thebibliography}

%

\begin{IEEEbiography}[{\includegraphics[width=1in,height=1.25in,clip,keepaspectratio]{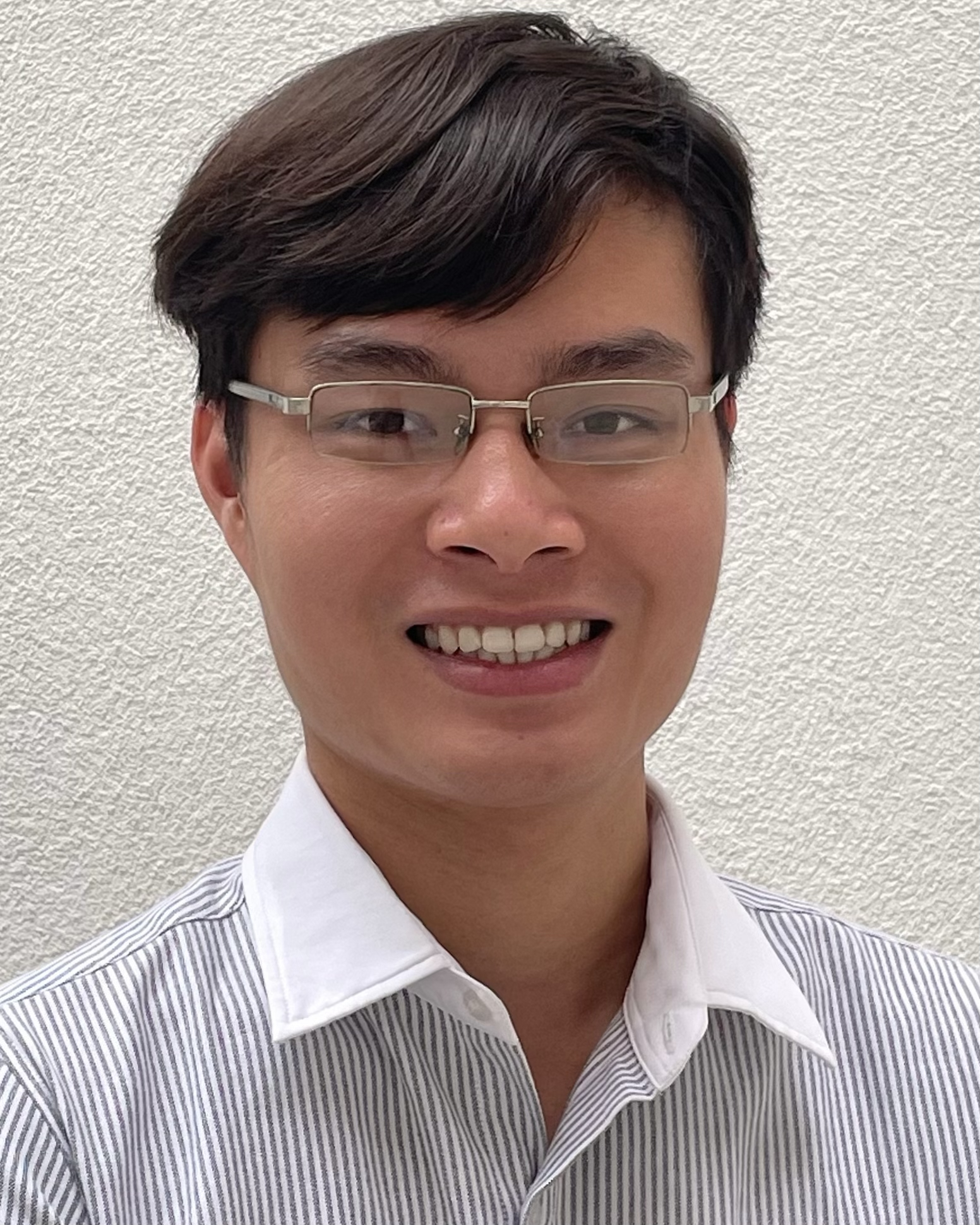}}]{Van Chien Le}(Member, IEEE)
received the M.Sc. degree in applied mathematics from Hanoi University of Science and Technology, Vietnam, in 2018, and the Ph.D. degree in mathematics from Ghent University, Belgium, in 2022.

Since 2022, he has been a Postdoctoral Researcher at the Department of Information Technology, Ghent University. His research interests include the numerical analysis of finite element and boundary element methods, with a particular focus on robust and stable frequency-domain and time-domain boundary integral equation solvers for electromagnetic scattering.

Dr. Le was awarded the IEEE Ulrich L. Rohde Innovative Conference Paper Award on Computational Techniques in Electromagnetics in 2023 and the URSI Young Scientist Award in 2026. 
\end{IEEEbiography}

\begin{IEEEbiography}[{\includegraphics[width=1in,height=1.25in,clip,keepaspectratio]{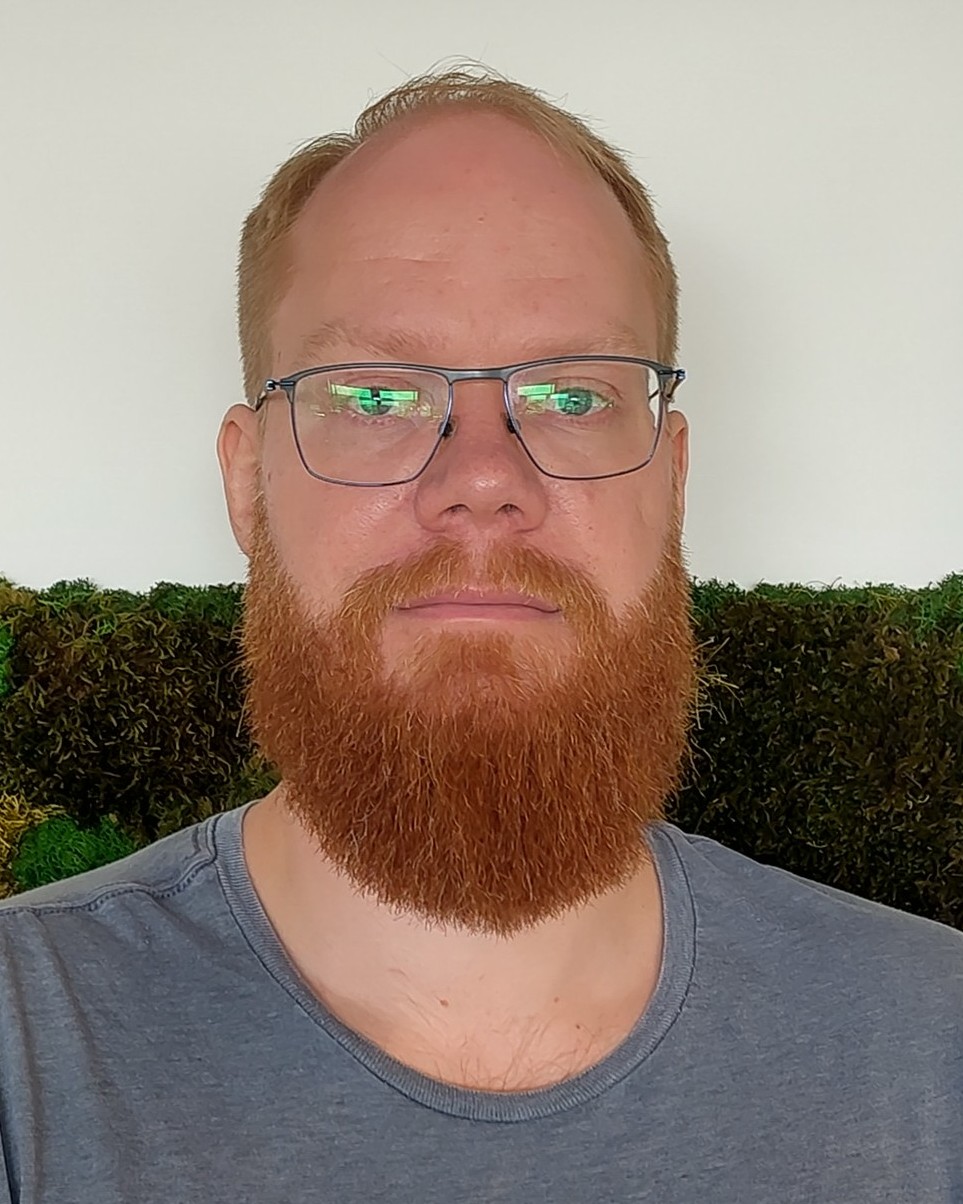}}]{Cedric M\"{u}nger}(Member, IEEE) received the M.Sc. degree in computational science and engineering from ETH Z\"urich, Switzerland, in 2020, and the Ph.D. degree in mathematical engineering from Ghent University, Belgium, in 2024. 

Since 2024, he has been a Postdoctoral Researcher in the Department of Information Technology at Ghent University. His research interests in computational electromagnetics include integral equation methods for composite structures, fast solvers in frequency-domain and time-domain, and their implementation on high-performance computing infrastructure. 

Dr. M\"unger received the Mini-Circuits Harvey Kaylie Best Paper Award at IEEE COMCAS in 2021.
\end{IEEEbiography}

\begin{IEEEbiography}
[{\includegraphics[width=1in,height=1.25in,clip,keepaspectratio]{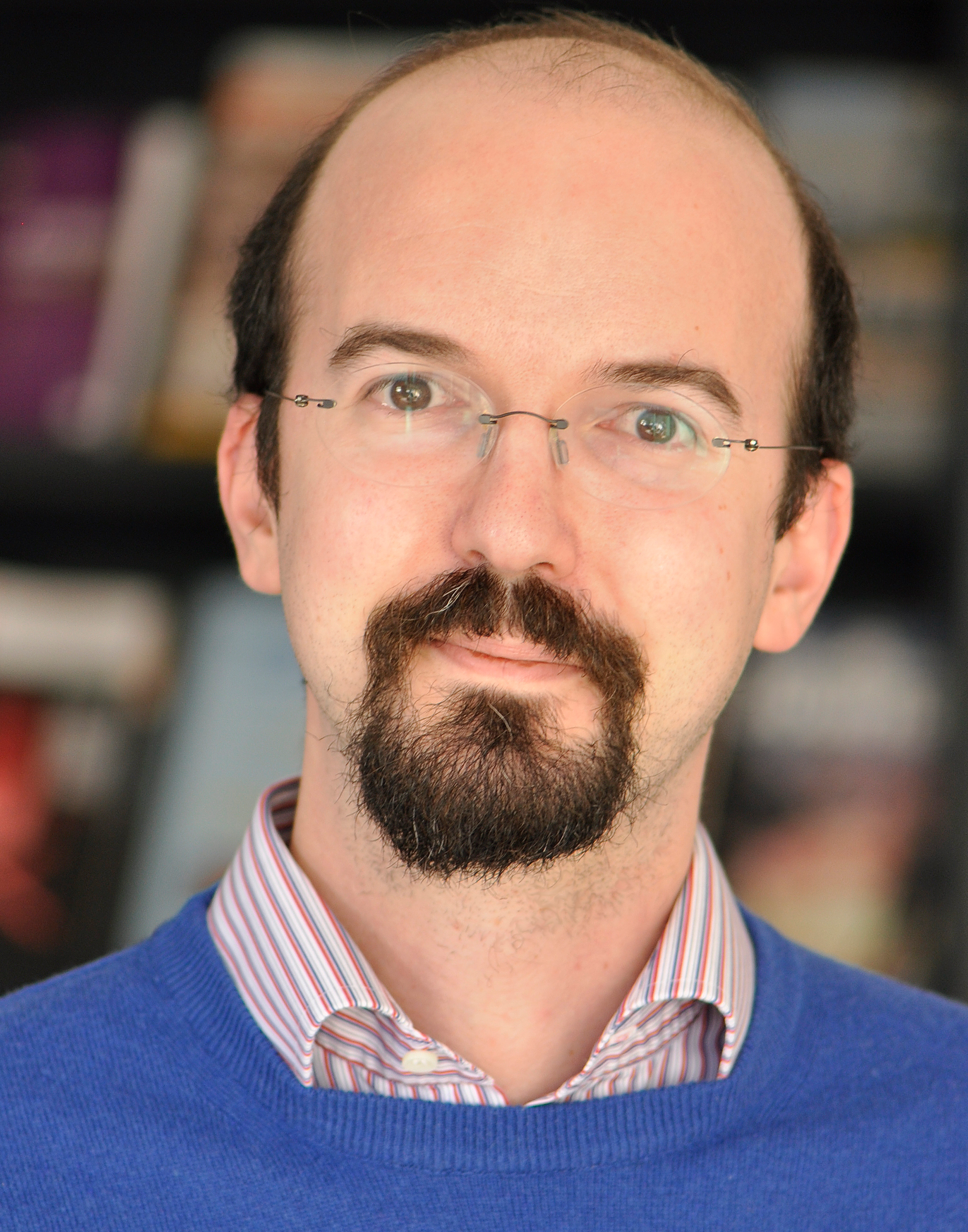}}]{Francesco P. Andriulli}(Fellow, IEEE)
received the Laurea in electrical engineering from the Politecnico di Torino, Italy, in 2004, the MSc in electrical engineering and computer science from the University of Illinois at Chicago in 2004, and the PhD in electrical engineering from the University of Michigan at Ann Arbor in 2008. From 2008 to 2010 he was a Research Associate with the Politecnico di Torino. From 2010 to 2017 he was an Associate Professor (2010-2014) and then Full Professor with the \'Ecole Nationale Sup\'erieure Mines-T\'el\'ecom Atlantique (IMT Atlantique), Brest, France. Since 2017 he has been a Full Professor with the Politecnico di Torino, Turin, Italy. His research interests are in computational electromagnetics including frequency- and time-domain integral equation solvers, well-conditioned formulations, fast solvers, low-frequency electromagnetic analyses, and modeling techniques for antennas, wireless components, microwave circuits, and biomedical applications with a special focus on brain imaging.

Prof. Andriulli received several best paper awards at conferences and symposia (URSI NA 2007, IEEE AP-S 2008, ICEAA IEEE-APWC 2015) also in co-authorship with his students and collaborators (EMTS 2025, ICEAA IEEE-APWC 2021, EMTS 2016, URSI-DE Meeting 2014, ICEAA 2009) with whom received also a second prize conference paper (URSI GASS 2014), a third prize conference paper (IEEE–APS 2018), seven honorable mention conference papers (ICEAA 2011, URSI/IEEE–APS 2013, 4 in URSI/IEEE–APS 2022, URSI/IEEE–APS 2023) and other three finalist conference papers (URSI/IEEE-APS 2012, URSI/IEEE-APS 2007, URSI/IEEE-APS 2006, URSI/IEEE–APS 2022). Moreover, he received the 2014 IEEE AP-S Donald G. Dudley Jr. Undergraduate Teaching Award, the triennium 2014-2016 URSI Issac Koga Gold Medal, and the 2015 L. B. Felsen Award for Excellence in Electrodynamics. 

Prof. Andriulli is a Fellow of the IEEE and of the International Union of Radio Science (URSI), and a member of Eta Kappa Nu, Tau Beta Pi, and Phi Kappa Phi. He serves as the 2026 President-Elect of the IEEE Antennas and Propagation Society and served as IEEE AP-S Vice-President of Publications 2025, as Editor-in-Chief of the \textit{IEEE Antennas and Propagation Magazine}, Track Editor for the \textit{IEEE Transactions on Antennas and Propagation}, and as an Associate Editor for the \textit{IEEE Antennas and Wireless Propagation Letters}, \textit{IEEE Access}, \textit{URSI Radio Science Letters}, and \textit{IET-MAP}.
\end{IEEEbiography}

\begin{IEEEbiography}
[{\includegraphics[width=1in,height=1.25in,clip,keepaspectratio]{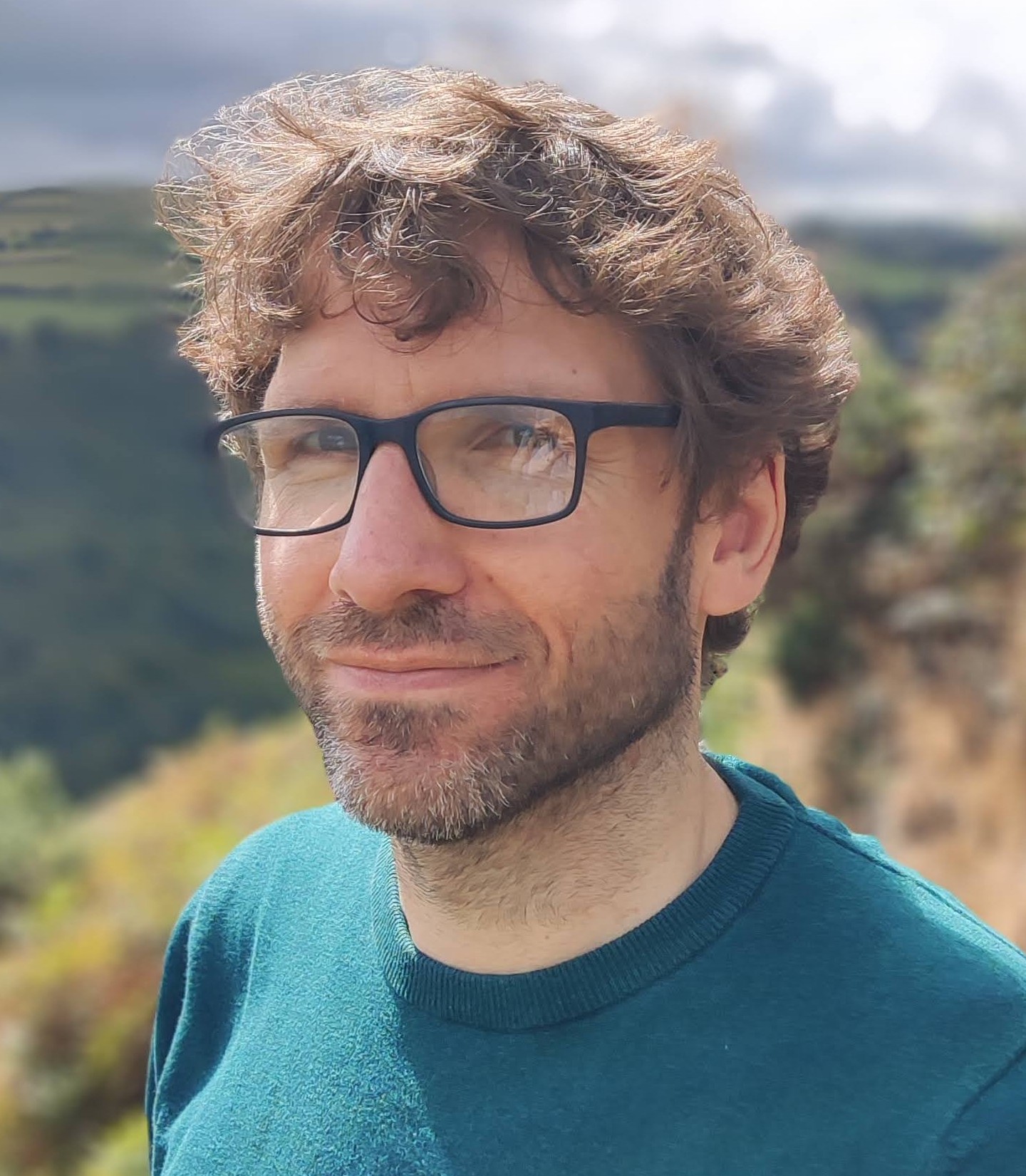}}]{Kristof Cools}(Member, IEEE) is full professor with the Department of Information Technology at Ghent University. Before joining Ghent University he has held positions at the TU Delft and the University of Nottingham. Kristof Cools has been awarded the ICEAA Young Scientist Award 2009, the URSI Young Scientist Award 2014, and the Rohde Most Innovative Conference Paper Award 2023. He has authored over 33 papers published in peer reviewed international journals and over 120 conference contributions, resulting in a Google Scholar recorded h-index of 19 and over 1800 citations. Dr. Cools has received funding from institutional, regional, industrial, national, and European funding bodies, including a Marie Skłodowska-Curie career integration grant and ERC consolidator/PoC project on boundary element methods for time-domain domain decomposition methods. 
\end{IEEEbiography}







\end{document}